\documentclass[aps,prl,reprint, superscriptaddress]{revtex4-1}
\usepackage{blindtext}
\usepackage{graphicx}
\usepackage{xcolor}
\usepackage{amsmath}
\usepackage{amsthm}
\usepackage{amssymb}
\usepackage{amsfonts}
\usepackage{comment}
\usepackage{siunitx}
\usepackage{xr}

\begin{document}
\title{Escape dynamics of active particles in multistable potentials}
\author{A. Militaru}
\thanks{These two authors contributed equally.}
\affiliation{Photonics Laboratory, ETH Zurich, CH-8093 Zurich, Switzerland}
\author{M. Innerbichler}
\thanks{These two authors contributed equally.}
\affiliation{Faculty of Physics, University of Vienna, 1090 Wien, Austria}
\author{M. Frimmer}
\affiliation{Photonics Laboratory, ETH Zurich, CH-8093 Zurich, Switzerland}
\author{F. Tebbenjohanns}
\affiliation{Photonics Laboratory, ETH Zurich, CH-8093 Zurich, Switzerland}
\author{L. Novotny}
\affiliation{Photonics Laboratory, ETH Zurich, CH-8093 Zurich, Switzerland}
\author{C. Dellago}
\thanks{e-mail: christoph.dellago@univie.ac.at}
\affiliation{Faculty of Physics, University of Vienna, 1090 Wien, Austria}

\makeatletter
\newcommand*{\addFileDependency}[1]{
  \typeout{(#1)}
  \@addtofilelist{#1}
  \IfFileExists{#1}{}{\typeout{No file #1.}}
}
\makeatother

\newcommand*{\myexternaldocument}[1]{%
    \externaldocument{#1}%
    \addFileDependency{#1.tex}%
    \addFileDependency{#1.aux}%
}

\maketitle

\textbf{Rare transitions between long-lived metastable states underlie a great variety of physical, chemical and biological processes. Our quantitative understanding of reactive mechanisms has been driven forward by the insights of transition state theory. In particular, the dynamic framework developed by Kramers marks an outstanding milestone for the field. Its predictions, however, do not apply to systems driven by non-conservative forces or correlated noise histories. An important class of such systems are active particles, prominent in both biology and nanotechnology. 
Here, we trap a silica nanoparticle in a bistable potential. To emulate an active particle, we subject the particle to an engineered external force that mimics self-propulsion. We investigate the active particle's transition rate between metastable states as a function of friction and correlation time of the active force. 
Our experiments reveal the existence of an optimal correlation time where the transition rate is maximized. This novel \emph{active turnover} is reminiscent of the much celebrated Kramers turnover despite its fundamentally different origin. Our observations are quantitatively supported by a theoretical analysis of a one-dimensional model. Besides providing a deeper understanding of the escape dynamics of active particles in multistable potentials, our work establishes a new, versatile experimental platform to study particle dynamics in non-equilibrium settings.
}

Transitions between long lives states are important for the understanding of chemical reactions \cite{Arrhenius1889a, VantHoff2010}, transitions between bistable configurations \cite{Ricci2017, Landauer1961b}, protein folding \cite{Sali1994, Frauenfelder1991}, motion of ligands in proteins \cite{Beece1980}, diffusion in solids through different domains \cite{DAgliano1975}, nuclear fission \cite{Bohr1939} and current switching in Josephson junctions \cite{Silvestrini1988}.
The transition rate, also called reaction rate, represents a central measure in this context, quantifying the frequency of transitions unfolding in meta- and multistable systems.

The first steps towards a quantitative understanding of reaction rates date back to the nineteenth century \cite{Arrhenius1889a, VantHoff2010, Eyring1935, Hanggi1990}, yet it was only in 1940 that Kramers developed the dynamic framework \cite{Kramers1940} widely used to this day. He considered a Brownian particle moving in a bistable potential and derived limiting expressions for high and low friction. Kramers realized that the transition rate constant disappears in both friction limits, and thus inferred the existence of a global maximum at some intermediate value of the damping, an aspect known today as the \emph{Kramers turnover} \cite{Grabert1988, McCann1999, Turlot1989, Troe1986}. It was only recently that the Kramers turnover has been measured in a single experimental system \cite{Rondin2017}.

Kramers' framework and its extensions \cite{Hanggi1982a, Pollak1986, Pollak1989, Pollak2013}, however, are a result of equilibrium dynamics and thus no longer apply in the presence of non-conservative forces. 
A particularly interesting example in which such forces are important is active matter. In active matter, the constituents draw on internally stored or externally supplied energy to propel themselves and drive the system out of equilibrium. Self-propulsion gives rise to various intriguing collective phenomena, such as swarming and orientation phase transitions \cite{Bechinger2016}. Even on an individual particle basis, self-propulsion holds great potential for applications in microscopic transport and sensing.
The perhaps simplest model for active matter is the \emph{active particle}---a particle subjected to thermal noise, dissipation, and to a self-propelling force of constant magnitude and Brownian orientation \cite{Volpe2014, Howse2007, Romanczuk2012}. These self-propelling agents arise in various contexts such as Janus particles \cite{Howse2007, Ghosh2013}, micro- and nanorobots \cite{Golestanian2007, Sundararajan2008}, motion of bacteria \cite{Berg2005, Gejji2012}, and active transport of biological macromolecules \cite{Brenner1990, Bressloff2013, Ghosh2013}. Understanding and controlling active particles represents thus a challenge of great importance in nanotechnology and medical sciences \cite{Ebbens2010}.

Previous attempts towards investigating the transition rates of active matter, crucial for their transport through constrictions and interfaces, are constrained to overdamped dynamics or to an activity induced by a velocity-dependent damping \cite{Geiseler2016, Schaar2015, Pohlmann1997, Burada2012}. Yet any type of movement in low-density media, for instance dilute gases, is heavily affected by inertial effects \cite{Scholz2018}. The Kramers turnover in particular represents an interesting example of inertial effects on transition phenomena. Furthermore, advancements in nanotechnology require the examination of automated and stochastic self-propulsion in various environmental conditions. The transition rate of active particles in the underdamped regime is thus a key question which has remained surprisingly unexplored to date. 

In this work, we experimentally investigate and theoretically analyze the transition rate of an active particle in a bistable potential over a wide range of frictions. We implement an active particle by applying an engineered stochastic force to an optically levitated nanoparticle. Our setup allows us to span both the overdamped and the underdamped motional regimes. We observe a new turnover as a function of the decorrelation time of the propulsion's orientation. The new turnover is of a different nature from its passive counterpart, and the two are shown to coexist in a two-dimensional parameter space. The experimental observations are in quantitative agreement with theoretical results and numerical simulations.

\begin{figure}[t!]
	\centering
	\includegraphics[width=8.6cm]{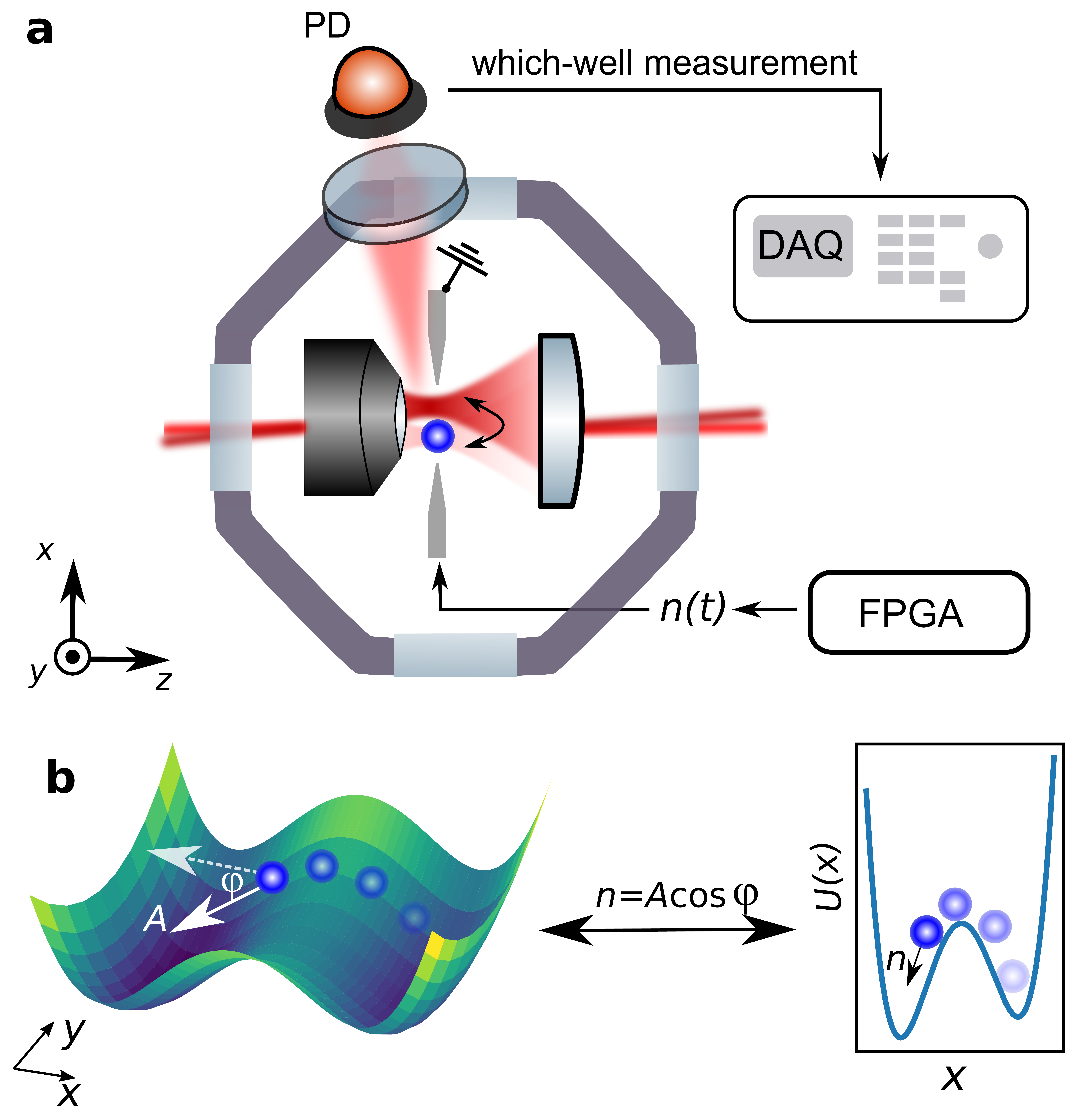}
	\caption{\textbf{Model under study and experimental setup. a,} Experimental setup. We create a bistable optical potential for a charged silica nanoparticle by focusing two cross-polarized beams through a 0.80~NA objective. We implement a \emph{which-well measurement} by collecting the laterally scattered light with a photodiode (PD). Owing to the different polarizations in the two wells, the intensity measured by the photodiode is a binary signal that indicates in which potential well the particle resides. The particle carries a finite net electric charge and is made active along $x$ with an electrostatic force generated by a voltage applied across lateral electrodes. This voltage is generated by a field-programmable gate array (FPGA). \textbf{b,} A two-dimensional active particle in a potential landscape that is bistable along the $x$-direction and harmonic along $y$. The active particle is propelled by a force of constant magnitude $A$ and Brownian orientation $\varphi$. The dimensions $x$ and $y$ are decoupled and the $x$ motion is equivalent to a particle moving in a one-dimensional bistable potential under a time-dependent force $n(t) = A\cos \varphi(t)$.}
	\label{fig:setup}
\end{figure}

\section{Experimental System}
The experimental setup is shown in Fig.~\ref{fig:setup}a. A  charged silica nanoparticle of nominal diameter $\SI{136}{nm}$ is trapped in a bistable optical potential. 
The bistable potential is realized by focusing two cross-polarized and frequency-shifted Gaussian beams through a high NA objective (wavelength $\lambda= \SI{1064}{nm}$). The two foci lie along the $x$-axis and their distance is controlled by a careful alignment of the relative angle between the beams. To implement a direct which-well measurement, we introduce a photodetector to monitor the light scattered by the particle in a direction perpendicular to the optical axis. Owing to the mutually orthogonal polarization of the two beams, together with the radiation pattern of a linear dipole (which emits no radiation along its axis), the recorded signal displays jumps as the particle transitions from one well to the other. The traces recorded by the photodetector are used for the study of the transition rates presented throughout this work. 

We apply an external electrostatic force that mimics the behaviour of active propulsion parallel to the potential's bistability direction. The voltage signal used to apply the active force is generated by a field-programmable gate array (FPGA), see Methods. Throughout this work, we study a particle actively propelled by a force of constant magnitude $A$, called activity, with stochastically changing direction $\varphi$. After projecting the active force onto the bistability direction $x$, the motion of the particle is described by the following Langevin equations:
\begin{subequations}
	\label{eq: equation of motion}
	\begin{align}
		&m\ddot{x} +m\Gamma_0 \dot{x} +\partial_x U(x) = A\cos\varphi + \mathcal{F}_\text{th}, \label{eq: COM} \\
		&\dot{\varphi} = \sqrt{2D_R}\eta_R (t). \label{eq:rotation}
	\end{align}
\end{subequations}
The position $x$ evolves in time under the influence of a frictional force proportional to the damping coefficient $\Gamma_0$, a conservative trapping force arising from the bistable potential $U(x)$, and thermal noise at temperature $T$ related to the friction via the fluctuation-dissipation theorem $\mathcal{F}_\text{th}=\sqrt{2m\Gamma_0 k_B T}\eta_\text{th} (t)$ \cite{Kubo1966}. The two mutually uncorrelated white noises $\eta_\text{th}$ and $\eta_R$ individually satisfy the properties $\left\langle \eta(t) \right\rangle =0$ and $\left\langle \eta(t)\eta(t')\right\rangle =\delta(t-t')$. The damping coefficient $\Gamma_0$ can be tuned by changing the pressure of the vacuum chamber \cite{Beresnev1990}. The orientation of the active force follows overdamped and purely diffusive dynamics associated with the rotational diffusivity $D_R$. In the following, we use $n = A\cos{\varphi}$ to refer to the one dimensional active force. The system is illustrated schematically in Fig.~\ref{fig:setup}b. 

\begin{figure}[t]
	\centering
	\includegraphics[width=8.6cm]{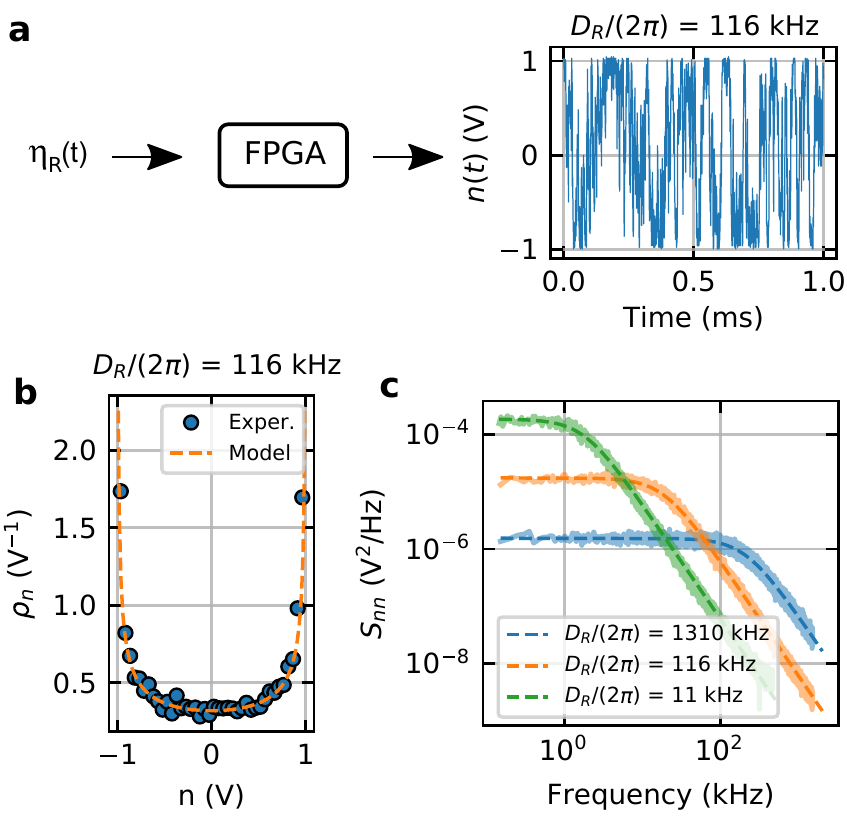}
	\caption{\textbf{Statistical properties of the active force.} The values shown are electric signals which act on the particle through the Coulomb force. \textbf{a,} Transformation of white Gaussian noise through an FPGA into the active force $n=A\cos\varphi$ of equation~\eqref{eq: equation of motion} ($D_R = 2\pi\times \SI{116}{kHz}$). \textbf{b,} Probability density of the realized active force in \textbf{a}. The arcsine distribution (dashed line) is expected from the projection onto one dimension of a constant force with fluctuating direction. \textbf{c,} Example spectra of active forces for three different values of $D_R$. Panels \textbf{b} and \textbf{c} highlight the non-Markovian and non-Gaussian nature of active propulsion. We refer to the Supplementary Information for a derivation of the theoretical curves shown as dashed lines.}
	\label{fig: active force}
\end{figure}

Figure~\ref{fig: active force} shows the measured characteristics of the active force, i.e., of the voltage output by our custom-programmed FPGA (see Supplementary Material).
 Specifically, in Fig.~\ref{fig: active force}a we show an example time trace of the active force for $D_R = 2\pi\times \SI{116}{kHz}$. Figure~\ref{fig: active force}b depicts the histogram of the trace in Fig.~\ref{fig: active force}a, and \ref{fig: active force}c shows three examples of the power spectral density (PSD) $S_{nn}$ for different rotational diffusivities. In stark contrast to the thermal fluctuations induced by the surrounding gas, the activity's noise history is non-Gaussian and coloured. 

\begin{figure*}[t]
	\centering
	\includegraphics[width=17.8cm]{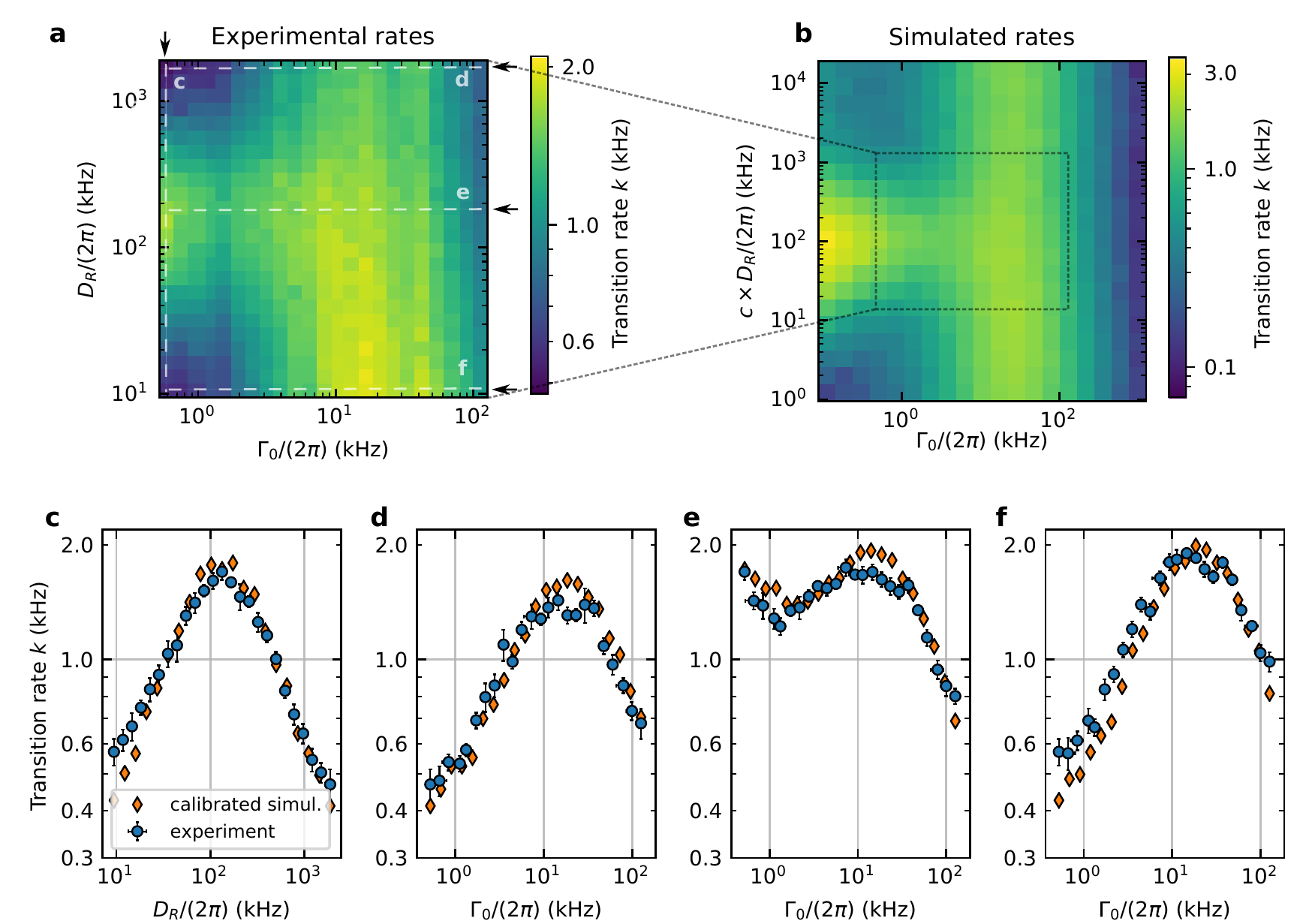}
	\caption{\textbf{Experimental and numerical transition rates. }\textbf{a,} Experimental transition rate constant $k$ as a function of rotational diffusivity $D_R$ and translational damping $\Gamma_0$. The dashed lines parallel to the arrows refer to the four cuts shown in \textbf{c, d, e, f}. At low $\Gamma_0$ we observe the novel active turnover along the $D_R$ axis. For fixed $D_R$ we recover a Kramers-like turnover. \textbf{b,} Computationally obtained transition rate constant (activity $A= \SI{9.35}{fN}$). The simulated $D_R$-axis is rescaled by the calibration factor $c = 2.24$, as described in the main text. The transition rate constant agrees within roughly $10\%$ with the calibrated numerical estimates. \textbf{c,} Active turnover. Vertical cut at $\Gamma_0 = 2\pi\times \SI{523}{Hz}$ of the experimental (blue circles) and of the simulated (orange diamonds) transition landscapes. \textbf{d,} Horizontal cut at $D_R = 2\pi\times \SI{1.8}{MHz}$. High rotational diffusivity in conjunction with sufficiently high friction leads to a recovery of inactive dynamics and of the Kramers turnover. \textbf{e,} Horizontal cut at $D_R = 2\pi\times \SI{166}{kHz}$, roughly the position of the active turnover. Starting at high $\Gamma_0$ and moving towards lower values, we initially encounter the Kramers-like turnover mainly arising from thermal transitions. Passing the local minimum, a further steady increase in transition rate constant is induced by the active propulsion facilitated by a weaker damping. \textbf{f,} Horizontal cut at $D_R = 2\pi\times \SI{9}{kHz}$. As $D_R$ approaches zero the active force becomes a modification to the potential with an effectively constant $\varphi$. The resulting Kramers-like turnover results from an average over the corresponding rate constants. Rate extraction and error bars are described in the Methods.} 
	\label{fig:turnovers}
\end{figure*}

\section{Results}
The central quantity of interest in the present study is the transition rate constant $k$, i.e., the typical frequency of transitions between the metastable states of the potential. It is extracted from the decaying autocorrelation of the which-well measurement (see Methods).
Figure~\ref{fig:turnovers}a showcases the transition rate constant as a function of rotational diffusivity $D_R$ and translational damping $\Gamma_0$. Each data point stems from a \SI{30}{s} long trajectory with fixed pressure and rotational diffusivity. We observe two perpendicular lines of cross-section maxima. 
The vertical line of maxima appears at roughly $\Gamma_0 = 2\pi\times \SI{20}{kHz}$ and corresponds to the Kramers turnover \cite{Rondin2017}. The second, horizontal one emerges at $D_R = 2\pi\times\SI{166}{kHz}$ and represents the central result of this work: an activity-induced turnover. 
We additionally depict four cross-sections highlighting the active turnover at $\Gamma_0= 2\pi \times \SI{523}{Hz}$ (Fig.~\ref{fig:turnovers}c), its passive Kramers counterpart at $D_R= 2\pi \times \SI{1.8}{MHz}$ (Fig.~\ref{fig:turnovers}d), a cut along the rotational diffusivity $D_R = 2\pi \times \SI{166}{kHz}$ that corresponds to the active turnover (Fig.~\ref{fig:turnovers}e), and the Kramers-like turnover at $D_R= 2\pi \times \SI{9}{kHz}$ (Fig.~\ref{fig:turnovers}f). 
The prominence of the active turnover decreases with increasing damping $\Gamma_0$, blends into the Kramers turnover and vanishes for very high dampings. 

In order to shed light on the nature of the active turnover, we implemented a numerical reconstruction of the observed transition rate landscape. The numerical reconstruction, displayed in Fig.~\ref{fig:turnovers}b, aims at recovering the landscape's key features using three fit parameters: barrier height, activity, and distance between potential wells.
The numerical reconstruction is in quantitative agreement with the experimental data throughout the observed parameter space. In addition to the three fit parameters, we introduce a calibration factor $c$ for the rotational diffusivity, needed to bridge the one-dimensional simulation and the more complicated three-dimensional reality:
Generally, in higher-dimensional systems it is possible for the particle to prefer different transition channels when driven by the activity or by thermal fluctuations. 
Such a multiplicity cannot occur in the one-dimensional model of equation~\eqref{eq: equation of motion}, 
it can however easily appear in the experiment, caused for instance by a slight misalignment between potential minima and activity axes. Additional details about the numerical reconstruction can be found in the Methods and Supplementary Material.

\section{Discussion}
The structure of the observed rate landscape in Fig.~\ref{fig:turnovers} is surprisingly rich. Nonetheless, its individual features can be understood on the basis of a few intuitive arguments stemming from a more rigorous analysis presented in the Supplementary Material. The existence of a maximum along the $\Gamma_0$-axis is analogous to the Kramers turnover in passive systems. Its renewed emergence supports the continued validity of Kramer's predictions even in non-conservative setups, extending their range of applicability. Next, we focus on the novel aspects arising specifically due to the presence of active propulsion. 
With the angle evolving according to Brownian motion, equation~\eqref{eq: equation of motion}, active propulsion can be interpreted as a stochastic force characterized by an exponentially decaying autocorrelation $\left\langle \cos \varphi(t) \cos \varphi(t') \right\rangle = \exp(-D_R |t-t'|)/2$. 
This relation identifies the rotational diffusivity as a measure of the characteristic timescale during which the active force's orientation does not change appreciably: the persistence time $\tau_A = D_R^{-1}$. This property is the key to reveal the mechanisms underlying the active turnover. To this end, let us conduct a simple thought experiment.

We start from the case of low rotational diffusivity: if the particle's orientation persists for a very long time, it is possible to consider the barrier to be modified by an additional tilt to the potential of $-Ax\cos \varphi$ (with constant $\varphi$). This modification to the barrier height facilitates the transition in one direction while hampering the reverse one. This trend becomes more apparent as $A$ increases, and at some point we would almost necessarily have to wait for the active force to change its sign to observe the next transition. For very large $A$ and small $D_R$ we expect to observe a single transition at best and remain stuck in the temporarily (or practically eternally) favoured well. The rate constant would need to effectively vanish in this extreme scenario.

Very high values of $D_R$ affect the system in a fundamentally different manner: if the activity's direction becomes completely decorrelated on the timescale of small positional displacements, its presence only increases the translational diffusivity in magnitude. In other words, fast rotating active forces raise the particle's effective temperature.
Over the course of a barrier transition the orientation $\varphi$ assumes all its possible realizations and can no longer persistently push the particle over the barrier. The transition rate constant is nevertheless enhanced given that larger diffusivities allow to scale barriers more easily in general. 
Specifically, as $D_R$ is decreased from infinity this effect becomes gradually stronger since the random mean square displacements caused by the activity increase in size on average. This implies a higher effective temperature at lower $D_R$ and subsequently higher transition rate constants $k$. The extreme scenario of infinitely fast rotation, on the other hand, will result in no net change of the effective translational diffusion and a recovery of inactive dynamics.
Finally, it is reasonable to expect a maximum in the transition rate constant if the persistence time is similar to the characteristic duration necessary to transverse the barrier, namely the average transition path time.
Then one typically retains the active force's aid during the whole transition, without inhibiting the reverse reaction any longer than necessary. 
Incidentally, this line of thinking closely follows Kramers' original argument for the existence of the passive turnover as a consequence of two opposing monotonic trends. Here, the position of the maximum is closely tied to the average transition path time as well, roughly emerging when the energy dissipated during the transition is comparable to the thermal energy $k_{\text{B}}T$.
The particle's trajectory then experiences thermal decorrelation that is sufficiently fast to avoid subsequent recrossings (which do not contribute to the rate) without losing the benefit of a more persistent direction of the velocity. 

With the timescales of energy dissipation and orientation decorrelation determined by separate parameters, the location of the passive and active turnover on their respective axis become practically independent from changes in the remaining variable, i.e., $D_R$ or $\Gamma_0$, respectively. This results in the cross-section maxima of Fig.~\ref{fig:turnovers} forming two approximately perpendicular lines. 
The last novel prominent feature, the turnover's gradual disappearance at large $\Gamma_0$, constitutes the property easiest to explain. Any sufficiently high friction can serve to slow down and practically arrest the particle's movement, leading to the rate constant's general decrease as one steps farther into the overdamped regime. Conversely, raising $A$ allows the active noise to compete against strong dissipative forces, making the phenomenon relevant for overdamped dynamics as well.

\section{Conclusion}
We have extended the analysis of the kinetics of transitions in a bistable system to active particles covering both the regimes of overdamped and underdamped motion. We have observed for the first time a novel turnover that arises as the persistence time of the active particle gradually increases from values much shorter to values much longer than the dynamical timescales of the system. A simplified one-dimensional description is sufficient to replicate the rich phenomenology of our experimental findings, and is capable of generating quantitatively consistent predictions. A full closed-form description that bridges the low with the high rotational diffusivity limits remains an open question tied to the complexity of the underlying Fokker-Planck equation. 

Our experimental platform can be adapted in the future to address various theoretical concepts arising in stochastic thermodynamics involving non-equilibrium systems and correlated noise histories. For instance, with the inclusion of a linear position measurement across a wide region of space, we can extend the model to a full three-dimensional activity rather than a one-dimensional projection. This setup would enable a rigorous investigation on the statistics of transition path times and the emergence of different pathways preferred by active and thermal transitions as a function of the activity. With the inclusion of a position feedback we can generate more complex kinds of position- and time-dependent activities, which have recently been shown to have intriguing tactic properties \cite{Geiseler2017}.
Furthermore, our externally applied stochastic force is not inherently constrained to mimic active propulsion, but rather allows for the introduction of any desired noise history into the system. This versatility opens up experimental opportunities in the field not tied to activity, such as fluctuation theorems in the presence of coloured noise.

\begin{footnotesize}

\section{Methods}
\textbf{Experimental setup.} The frequency shift (\SI{80}{MHz}) between the two beams is introduced by an acousto-optic modulator (AOM) acting on a $\lambda= \SI{1064}{nm}$ wavelength laser. The AOM also controls the power of the two beams, equal to $70\pm1$ ($110\pm1$)~mW for the $x$- ($y$-) polarized beam. After the focus, the beams are recollimated and separated with a polarizing beam splitter (PBS). We use a standard homodyne position measurement on the $x$-polarized beam to characterize the potential curvatures \cite{Gieseler2012}. The curvatures are estimated through time-frequency analysis: the estimated values are $\omega_1 = 2\pi\times(73.0\pm 0.5)$~kHz for the $x$-polarized well and $\omega_2 = 2\pi\times(82\pm 4)$~kHz for the $y$-polarized well. 
The proportionality coefficient between damping and pressure is inferred via the power spectral density (PSD) of the particle when the $y$-polarized laser is turned off \cite{Hebestreit2018a}. Specifically, the PSD of a harmonic oscillator is given by a Lorentzian whose width equals the damping coefficient.\\

\textbf{Active force.} The active force is realized with a field-programmable gate array (FPGA) that takes as input a white noise created by a function generator. The FPGA integrates equation~\eqref{eq:rotation} to generate the active force as output. We exploit the net charge carried by the particle and apply the active force electrostatically through two electrodes mounted along the bistability direction \cite{Frimmer2017d}. The experimental value of the activity can be determined from the response of the particle to a known modulated voltage. Within 50\% accuracy due to the mass uncertainty, we estimated $A=\SI{6.8}{fN}$. A potential barrier height $\Delta U$ of a few $k_\text{B} T$ and diffraction limited width $\Delta x \approx \lambda/2$ gives rise to conservative forces with a typical magnitude of $\Delta U/\Delta x \approx \SI{8}{fN}$. In order to induce a measurable effect the activity needs to be of the same order of magnitude, as is the case here.\\

\textbf{Transition rate estimation.} Rate coefficients appear within the framework of the eponymous rate equations. This type of differential equation aims to describe the evolution of local concentrations or population fractions in a set of states $\{A,B,\dots\}$ subjected to reactive transitions. 
Our case, a bistable system with population fractions $c_A$ and $c_B$, represents a particularly simple example described by $\dot{c}_A = -k_{AB}c_A + k_{BA}c_B = k(c_{A,eq}-c_A)$. With the asymmetry between rate constants $k_{AB}$ and $k_{BA}$ attributable to the stationary concentration $c_{A,eq} =k_{BA}/(k_{AB}+k_{BA})$, the transition rate $k$ represents the dynamical parameter governing the speed of equilibration. Within the approximation of rate equations, $c_A(t)= (c_A(0)-c_{A,eq})e^{-kt}+ c_{A,eq}$ approaches its equilibrium value exponentially and the rate can be extracted from sufficiently long sample trajectories \cite{Rondin2017}. The rates in Fig.~\ref{fig:turnovers}, more specifically, are extracted from \SI{30}{s} long trajectories. We split each trajectory in ten segments and compute the average rate and its standard deviation.\\

\textbf{Numerical reconstruction.} The simulation results complementing our experimental findings stem from the optimization of an effective, one-dimensional potential along with the activity A. We use a bistable, piece-wise parabolic potential, continuous up to its first derivative and tune its barrier width/curvature $\omega_B$ and height $h$. Equation~\eqref{eq: equation of motion} is then numerically integrated for values of $\Gamma_0$ and $D_R$ spaced on a logarithmic grid. We employ the OVRVO integrator devised by Sivak et al.\ for the particle's propagation \cite{Sivak2014}. The curvatures of the well-parabolas respect the experimentally determined particle frequencies. The obtained transition landscapes are compared to the experimental reference w.r.t.\ a small number of effective quantities related to the active and passive turnover, respectively: maximum height, its position on the respective axis, and turnover width at half maximum, all evaluated on a logarithmic scale. We proceed to locate the parameter set of $h$, $\omega_B$, and $A$ that leads to the minimal deviation in the aforementioned measures on a discrete, linearly spaced grid of desired resolution. Starting from cautious a-priori estimates of these values, we rely on a bisection-like approach to locate said minimum. To simplify this procedure, note that at high $D_R$ one essentially recovers inactive dynamics, allowing us to optimize $h$ and $\omega_B$ independently of $A$. Additional details on the simulation and optimization procedure can be found in the Supplementary Material.

\end{footnotesize}

\section{Acknowledgements}

This research was supported by the Swiss National Science Foundation (grant no.\ 200021L\_169319 and no.\ 51NF40-160591) and the Austrian Science Fund (grant no.\ I3163-N36). The authors acknowledge P. Back, E. Bonvin, F. van der Laan, A. Nardi, J. Piotrowski, R. Reimann and D. Windey for fruitful discussions.

\bibliographystyle{apsrev4-1}
\bibliography{referencesShort} 

\end{document}


\title{Supplementary Information:\\ Escape dynamics of active particles in multistable potentials}
\author{A. Militaru}
\thanks{These two authors contributed equally.}
\affiliation{Photonics Laboratory, ETH Zurich, CH-8093 Zurich, Switzerland}
\author{M. Innerbichler}
\thanks{These two authors contributed equally.}
\affiliation{Faculty of Physics, University of Vienna, 1090 Wien, Austria}
\author{M. Frimmer}
\affiliation{Photonics Laboratory, ETH Zurich, CH-8093 Zurich, Switzerland}
\author{F. Tebbenjohanns}
\affiliation{Photonics Laboratory, ETH Zurich, CH-8093 Zurich, Switzerland}
\author{L. Novotny}
\affiliation{Photonics Laboratory, ETH Zurich, CH-8093 Zurich, Switzerland}
\author{C. Dellago}
\thanks{e-mail: christoph.dellago@univie.ac.at}
\affiliation{Faculty of Physics, University of Vienna, 1090 Wien, Austria}

\maketitle
\myexternaldocument{main}

\section{Extraction of Rate Coefficients}

Let us explicitly consider a two-state system characterized by the concentrations/population fractions $c_A$ and $c_B$. In the context of single particle systems one can define $c_A(t)$ via an ensemble of a large number of replicas. Then, $c_A(t)$ corresponds to the fraction of replicas in state $A$ at time $t$.
Subjected to reactive dynamics, $c_A$ evolves according to the rate equation
\begin{equation}
\dot{c}_A = -k_{AB}c_A + k_{BA}c_B = k_{BA}-(k_{AB}+k_{BA})c_A.
\label{eq:rateEquation}
\end{equation}
The rate constants $k_{AB}$ and $k_{BA}$ respectively prescribe the speed of the reaction from $A$ to $B$ and from $B$ to $A$.
Population/particle conservation enforces a constraint of the form $\sum_X c_X =1$, used above to arrive at the rightmost expression. It is straightforward to determine the solution as
\begin{equation}
c_A(t) = c_A(0)e^{-(k_{AB}+k_{BA})t}   + \frac{k_{BA}}{k_{AB}+k_{BA}}\left[1-e^{-(k_{AB}+k_{BA})t}  \right] .
\label{eq:rateEqSolv}
\end{equation} 
Hence, the populations approach their equilibrium values $c_{A,eq}$ and $c_{B,eq}$ exponentially with a decay-time controlled by the sum of the rate constants $k=k_{AB}+k_{BA}$.
We underline this point by reformulating the above equation in terms of the equilibrium/stationary state concentration $c_{A,eq}$
\begin{equation}
c_A(t) = \left[ c_A(0)-c_{A,eq} \right]  e^{-kt}  + c_{A,eq},
\end{equation} 
where the equilibrium population is given by $c_{A, eq}=k_{BA}/(k_{AB}+k_{BA})$.
The rate constants can therefore be extracted from the exponential decay/growth of the concentration. 
For the initial condition $c_A(0)=1$, the population $c_A(t)$ can be interpreted as the probability of finding the particle in well A at time t provided it was located there at time $t=0$. 
Contact between the phenomenological rate equation~\eqref{eq:rateEquation} and the experimental time traces of the system can be made by considering the indicator function $h_A$, which is a function equal to unity inside well $A$ and zero otherwise. Then, the average $\left\langle h_A(t) \right\rangle$ of the indicator function corresponds to the population $c_A(t)$.

In terms of the indicator function $h_A$, the conditional probability to find the system in $A$ at time $t$, provided it was there at time $t_0 = 0$, can be expressed using the time correlation function of the population as $\left\langle h_A(0) h_A(t) \right\rangle / \left\langle h_A \right\rangle$. Here, the average is taken over a long (equilibrium/stationary state) trajectory. Assuming that for sufficiently long times the kinetics of the system is described by the rate equation, one obtains
\begin{align}
\label{eq: autocorrelation fit}
\left\langle h_A(0) h_A(t) \right\rangle &= c_{A,eq}(1-c_{A,eq})e^{-kt}+c_{A,eq}^2, \\
\left\langle \Delta h_A(0) \Delta h_A(t) \right\rangle &= \left\langle \Delta h_A(0)^2 \right\rangle e^{-kt},
\end{align}
where $\Delta h_A$ designates $h_A-\left\langle h_A \right\rangle $ and we took advantage of the relations $\left\langle h_A \right\rangle = c_{A,eq}$ and $h_A^2= h_A$.
Consequently, one can extract the rate coefficient $k$ by performing a single parameter fit to the (normalized) autocorrelation function. In reality, however, the rate equation only represents a convenient simplification of the system at hand and deviations from purely exponential decays are expected, for instance due to short-time correlations. In order to avoid including short-time correlations in the rate estimations, we fit equation~\eqref{eq: autocorrelation fit} only over time scales where their effect has disappeared. 

\newcommand{\figref}[1]{{Fig.~#1}}
\newcommand{\kb}{k_{\text{B}} }
\newcommand{\eff}{_{\text{eff}} }

\section{Activity and Effective Barrier Height}

To supplement the experimental findings delineated in the main part of the document, we provide additional context by studying the transition dynamics theoretically and computationally. 
For clarity, we direct our focus mainly on the phenomenology
and select a particularly simple and frequently used form of the potential, i.e.\
\begin{equation}
\label{eq: quartic potential}
 U(x) = h\left(1-\frac{x^2}{\sigma^2} \right)^2.
\end{equation}
The parameter $h$ gives the height of the barrier and the minima of the symmetric potential are located at $\pm \sigma$. All quantities with dimensions of inverse time are subsequently given in terms of $k_0 = t_0^{-1}= \sqrt{\kb T/m \sigma^2}$. We investigate the shape of the resulting transition rate landscape as function of the translational damping $\Gamma$ and rotational diffusivity $D_R$. We generate and analyse individual long trajectories for each set of $\Gamma$ and $D_R$ on a logarithmically spaced grid, whilst all other parameters remain fixed. Further details about the simulation are delineated in Sec.~\ref{sec:simu}.

\begin{figure}	
    \centering
    \subfloat{\includegraphics[width=.45\linewidth]{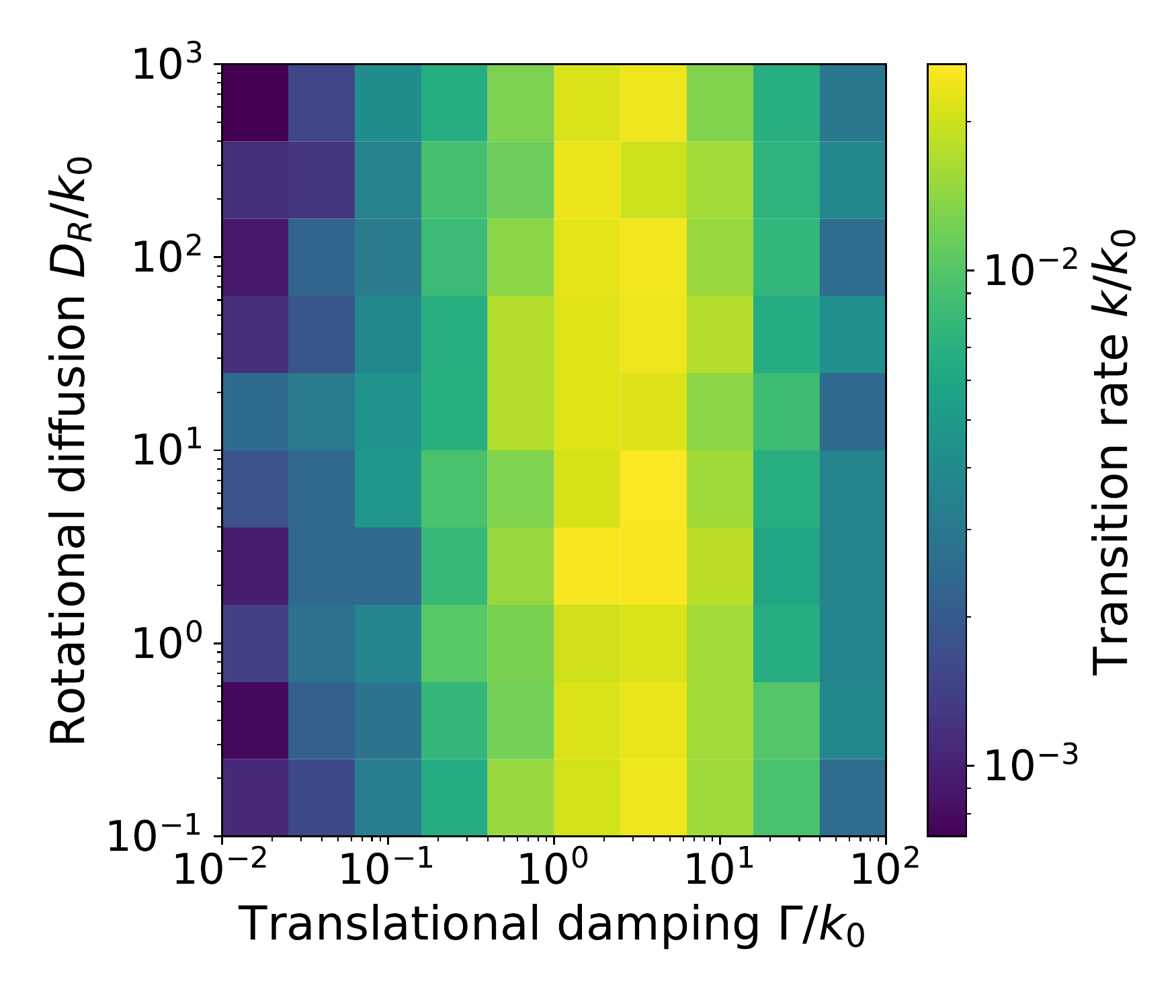}}
	\subfloat{	\includegraphics[width=.45\linewidth]{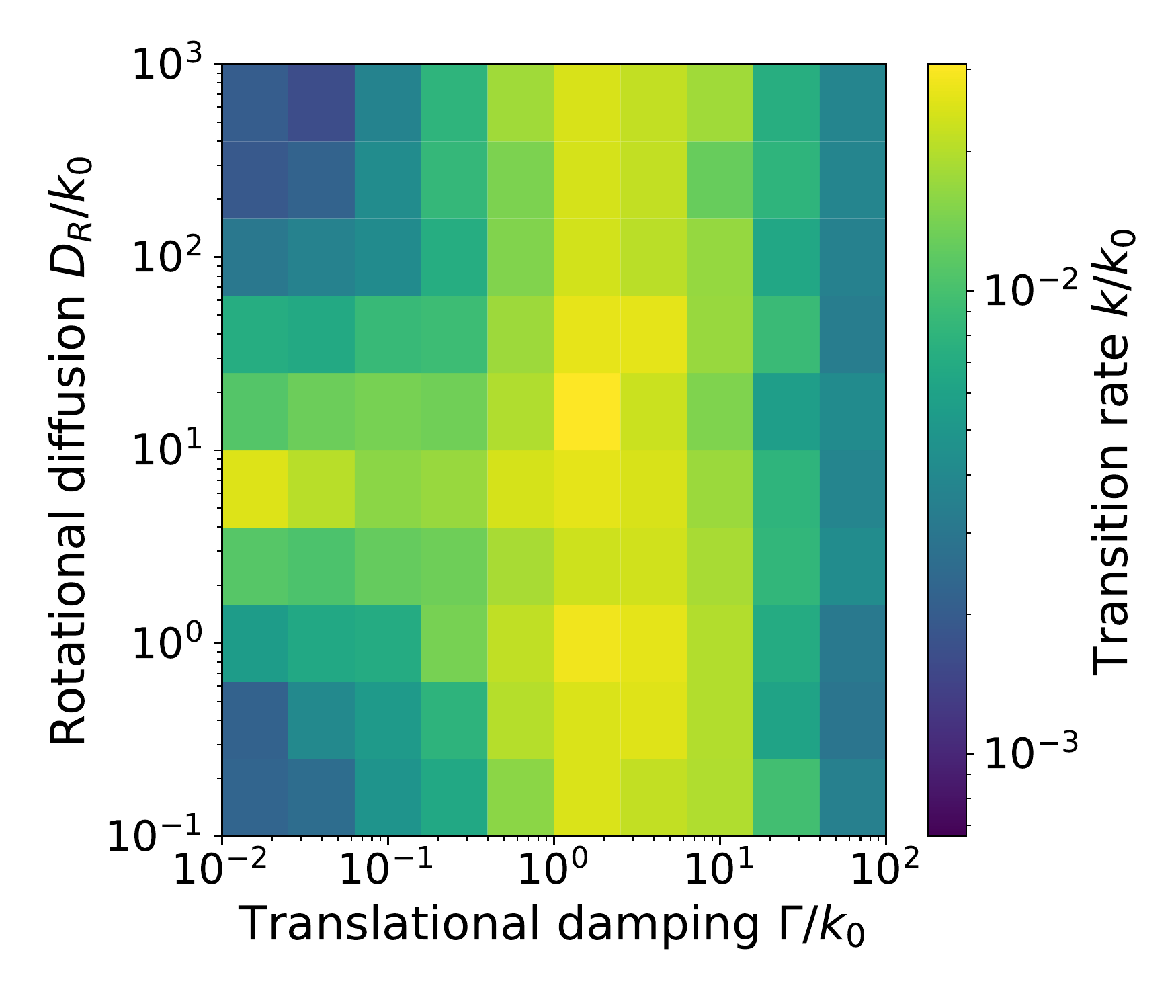}}	
	\\
	\subfloat{
	\includegraphics[width=.45\linewidth]{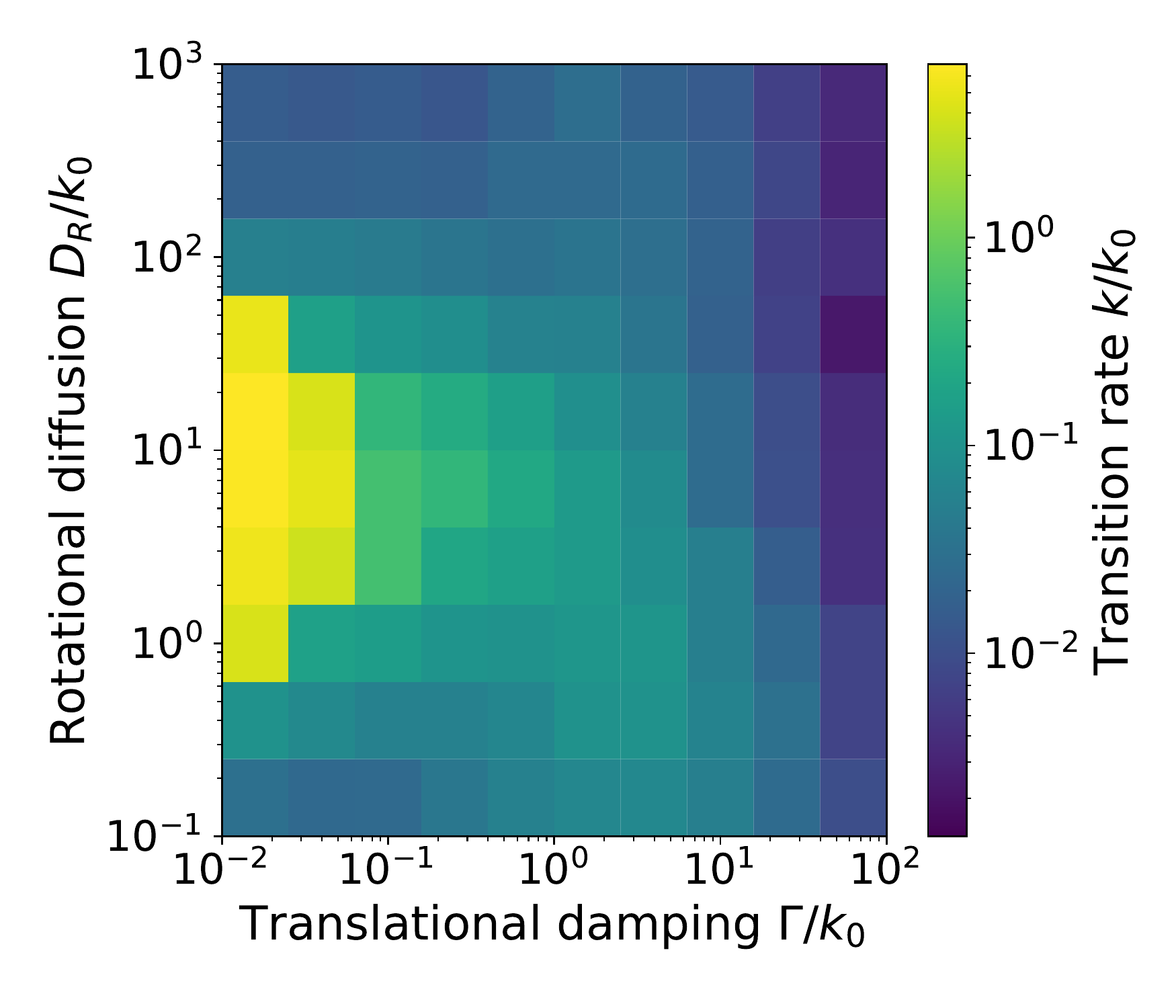}}
	\subfloat{
	\includegraphics[width=.45\linewidth]{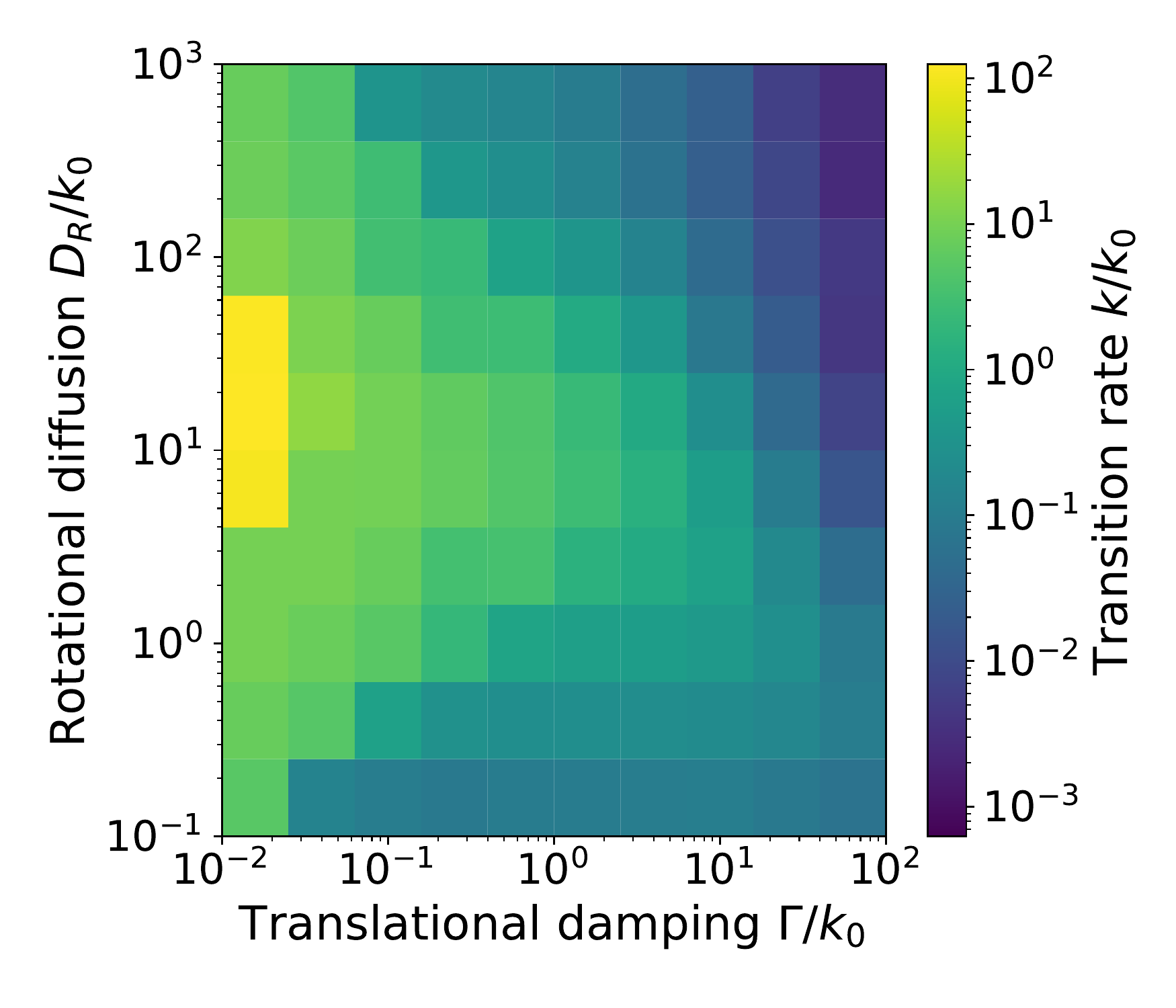}}
	\caption{Transition rate as a function of the translational damping and rotational diffusion for four activity magnitudes $A\sigma/\kb T=0.25,\, 1.00 ,\, 4.00 ,\, 16.0$. As $A$ increases the active turnover appears and becomes increasingly more prominent, whereas the Kramers turnover retains its magnitude. Simulation parameters: barrier height $h=4\kb T$, time step $\num{e-2}t_0$, trajectory length per data point $T=\num{2e+4}t_0$. }
	\label{fig:start}
\end{figure}

Some illustrative examples of the resulting rate landscapes with various activity values are depicted in \figref{\ref{fig:start}}.
Starting our discussion from low (almost vanishing) activity, one observes the standard Kramers turnover only. As one would expect in this case, the transition rate rises to its maximum at some moderate value of $\Gamma$ and remains virtually unaffected by the active contributions. Stronger activities, however, lead to data exhibiting some interesting features: a second line of maxima appears in the direction of the rotational diffusion axis $D_R$. This line seems to stop abruptly at the position of the standard Kramers turnover. Further increasing $A$ results in an evermore prominent active turnover, completely superseding its original counterpart at its horizontal position. The Kramers turnover has not disappeared, yet its magnitude is far too small to be seen on a linear scale. The newly found active turnover is unable to push past the limit imposed by the passive one and disappears gradually as the damping is increased. For even higher $A$, it becomes possible for the active turnover to stretch farther into the overdamped regime, however, simultaneously becoming more prominent at lower $\Gamma$ values. 

We have provided an intuitive motivation of the novel features arising due to active propulsion in the main text. In the following, we proceed to investigate the active turnover more quantitatively. Following Kramers' example, we demonstrate the existence of a turnover by examining the limits of high and low rotational diffusion individually. Instead of directly jumping to these limiting scenarios, however, it is instructive to discuss the activity's non-equilibrium nature.

Because it continuously feeds energy into the system, active propulsion constitutes an example of a non-conservative force. As such, the system's stationary state $p_s(x,v,\varphi)$ is not described by the Boltzmann distribution. The stationary state becomes explicitly dependent on the damping and rotational diffusion, distinctly setting the underlying physics apart from the properties generating the Kramers turnover: In conservative settings, the rate exhibits a maximum at moderate friction due to purely dynamical effects, while the probability density in phase space is independent of the damping $\Gamma$. 

\begin{figure}[tbp]
	\centering
    \subfloat{
	    \includegraphics[width=.45\linewidth]{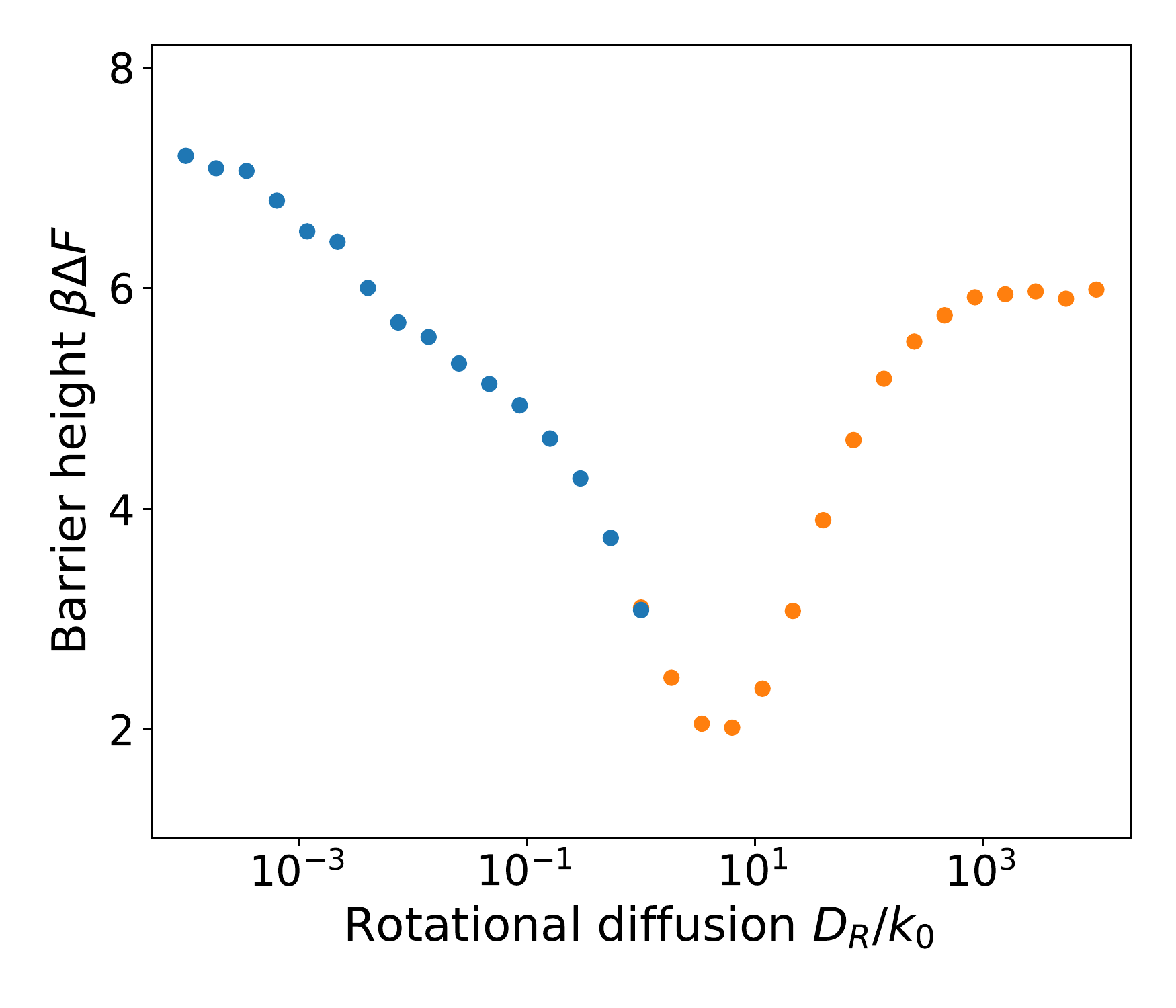}}
    \subfloat{
		\includegraphics[width=.45\linewidth]{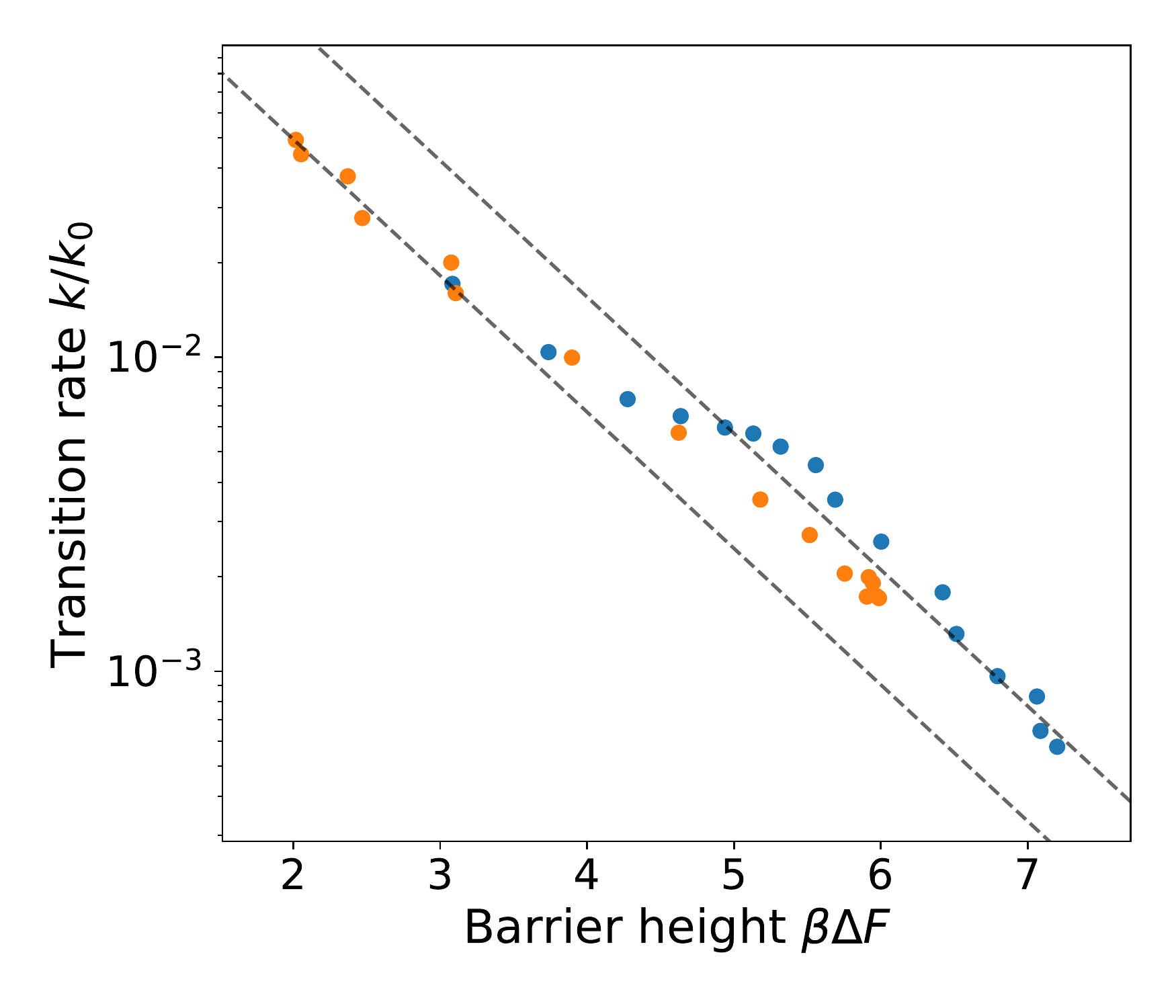}}
	\caption{Left: Free energy barrier height $\beta \Delta F$ as a function of rotational diffusivity $D_R$. Right: Transition rate $k$ as a function of free energy barrier height $\beta \Delta F$. The data points stem from trajectory simulations in the quartic potential of equation~\eqref{eq: quartic potential} with barrier height $h=6\kb T$, activity $A= 3\kb T/\sigma $ and at a damping of $\Gamma =0.2k_0$. The trajectory length per point equals $T=\num{e+6} t_0$. The showcased data encompass two separate sets of simulations for the high and low $D_R$ regime only differing in the size of time step $\Delta t$. 
	We used $\Delta t = 0.01 t_0$ for the low $D_R$ part (blue) and $\Delta t = 0.001 t_0$ for the fast rotation section (orange), as a finer discretization is needed to thoroughly resolve the evolution of the angle.
	The two gray lines accompanying the generalized Arrhenius plot indicate the slope expected theoretically. A switch in the rate's prefactor seems to emerge around $4\kb T$ - about a factor two in magnitude - not attributable to stationary effects. Nonetheless, the data's general consistency with the theoretical approximation points to the barrier height's central role in determining $k$.  }
	\label{fig:arrhenius}
\end{figure}

While usually introduced in equilibrium thermodynamics and statistical physics, it is useful to define the free energy $F$ for non-conservative settings analogously as the negative natural logarithm of the stationary state via $\beta F = -\ln p_s(x,v,\varphi)$ \cite{Sivak2012}. 
Adopting the view of transition state theory \cite{Dellago2009}, the rate depends exponentially on the height of the free energy barrier $\Delta F$, i.e.\ it is proportional to the probability to reside atop the barrier
\begin{equation}
	k\propto e^{-\beta \Delta F}.
	\label{eq:rateProp}
\end{equation}
Even small modifications of $\Delta F$ affect the rate landscape considerably. Studying the variation of $\Delta F$ as a function of $A$, $\Gamma$, and $D_R$ leads to a more quantitative understanding of the emergence of the active turnover. The Arrhenius plot depicted in \figref{\ref{fig:arrhenius}} serves to establish $\beta \Delta F$ as the key quantity in shaping the activity-dominated part of the rate landscape. The fact that only comparatively small deviations from the postulated exponential decay arise suggests that most relevant effects are indeed summarized into this single measure. This circumstance constitutes the basis of the fairly simple quantification approaches laid out below. 

Our physical model is described by a set of Langevin equations. Therefore the evolution of its phase space density $p(x,v,\varphi, t)$ is governed by the associated Fokker-Planck equation (we set the mass to unity for notational simplicity)
\begin{equation}
\label{eq: FPE}
\partial_t p (x,v,\varphi) = 
-v\partial_x p + \partial_v(\Gamma v+U'(x)-A\cos\varphi)p+ 
D\partial_{vv}^2 p +D_R\partial_{\varphi\varphi}^2 p. 
\end{equation}
The stationary distribution $p_s$ and free energy $F$ are in principle obtained by imposing $\partial_t p = 0$. Unfortunately, as of now a closed form solution of the resulting FPE remains elusive, forcing us to resort to numerical or approximate treatments. The upcoming derivations for the high and low $D_R$ limits are inherently based on the FPE, although some aspects are understood more easily in the Langevin picture. The theories successfully reproduce the transition rates obtained computationally, capturing the essential aspects to affect the free energy barrier height.

\section{High Rotational Diffusion Limit}

Let us first consider the limit of large $D_R$, in which the persistence time of the active force is far shorter than all remaining physical timescales. This effectively implies that before the particle changes its position in phase space by an appreciable amount, the direction of its active propulsion completely decorrelates, allowing it to perform several stochastic rotations on the spot. Therefore, at the timescale of physical motion 
the active noise behaves as if subsequent realizations were statistically independent, effectively putting it into the same memory class as white noise. Based on this reasoning, a modification of the diffusion constant or, equivalently, the introduction of an effective temperature provides an adequate approximation of the impact of the active force on the system. To justify this approximation quantitatively, let us quickly consider the limit $D_R \to \infty$. For an  orientation changing infinitely fast, the angle distribution must be uniform at each point in reduced phase space $(x, v)$. The effective dynamics in reduced phase space then results from replacing all $\varphi$-dependent terms with their local averages. 
The expectation value of the active force $A\cos\varphi$ vanishes given a uniformly distributed angle, leading us to the same equation of motion as in the absence of activity:
\begin{equation}
	\dot{v} = -\Gamma v -U'(x)+\sqrt{2m\Gamma \kb T}\eta(t).
\end{equation}
Next, we consider the equation of motion at the translational timescale for a finite (but still very high) rotational diffusion. We emphasise that for fast rotational diffusion it is possible to choose a duration $\Delta t$ that allows for substantial changes in $\varphi$ whilst keeping $x$ and $v$ effectively constant.
The deterministic displacement in $x$ and $v$ throughout the time step $\Delta t$ should then be virtually unaffected by the details of the noise history in-between. This allows us to investigate the displacements attributed to the (active) noise terms without the presence of any deterministic drifts due to potential forces and friction. Our argument should even work in the presence of multiplicative noise as the current phase point, for all intents and purposes, turns into a parameter. 

The total velocity displacement due to stochastic forces $\Delta_s v$ over $\Delta t$ can be expressed as
\begin{equation}
\label{eq: velocity change}
\Delta_s v = \int_{0}^{\Delta t} dt\left(  A\cos \varphi (t)+ \sqrt{2D}\eta(t)\right) .
\end{equation}
We consider velocity displacements because the active turnover resides primarily in the underdamped regime. The present argumentation can be applied to the overdamped approximation as well. We will do so when studying its disappearance for increasing $\Gamma$. We compute the mean square displacement associated with equation~\eqref{eq: velocity change} to obtain
\begin{align}
\left\langle \Delta_s v^2 \right\rangle  &= \left\langle\int_{0}^{\Delta t} dt \int_{0}^{\Delta t} dt' \left(  A\cos\varphi(t)+ \sqrt{2D}\eta(t) \right)\left(  A\cos\varphi(t')+ \sqrt{2D}\eta(t') \right) \right\rangle = \nonumber\\
& =  \int_{0}^{\Delta t} dt \int_{0}^{\Delta t} dt' A^2 \left\langle \cos \varphi(t) \cos \varphi(t') \right\rangle + 2D\left\langle \eta(t) \eta(t') \right\rangle = \nonumber\\
& =  \int_{0}^{\Delta t} dt \int_{0}^{\Delta t} dt'\left( \frac{A^2}{2} e^{-D_R |t-t'|} + 2D\delta(t-t') \right).
\end{align}
The expectation values of the mixed terms disappear due to statistical independence, splitting them into a product of expectation values equal to zero. The autocorrelation of the white noise yields a Dirac-delta by definition and the cosine term is derived in Sec.~\ref{sec: activity}. The final result reads
\begin{align}
\label{eq: displacement}
\left\langle \Delta_s v^2 \right\rangle  &= 2D\Delta t+ \frac{A^2}{D_R} \left(\Delta t + \frac{1}{D_R} \left( e^{-D_R \Delta t}-1 \right)\right) = \nonumber\\ &= 2\left(D+\frac{A^2}{2D_R}\right)\Delta t + \mathcal{O}(D_R^{-2}).
\end{align}
Comparing equation~\eqref{eq: displacement} with inactive dynamics, we obtain an effective translational diffusion equal to $D_\text{eff}=D+A^2/(2D_R)$, hereby neglecting the small terms of order $D_R^{-2}$. These higher order terms emerge from the fact that the active noise is inherently not a Gaussian white noise even in the limit considered. For instance, white noise can induce arbitrarily large displacements throughout $\Delta t$, whereas active noise has an upper bound in magnitude at $ A \Delta t$. Nonetheless, these differences become negligible for very high $D_R$. In terms of an effective temperature one obtains (now including the mass for completeness)
\begin{equation}
\label{eq: effective temperature}
D_{\text{eff}} = m\Gamma\kb T\eff = m\Gamma\kb T\left( 1+ \frac{A^2}{2m\Gamma \kb T D_R} \right). 
\end{equation}
Attributing all relevant modifications between the active and inactive stationary distribution to this effective temperature leads us to the relation
\begin{equation}
	\beta \Delta F =  \frac{\Delta U}{\kb T\eff} ,
\end{equation} 
where $\Delta U$ denotes the energy barrier of the trapping potential.

At this point we are able to approximate the high rotational diffusion part of the landscape with a single parameter fit of the form
\begin{equation}
k(D_R) = c_0 e^{-\Delta U/\kb T\eff(D_R)}.  
\end{equation}

\begin{figure}[tbp]
	\centering
	\subfloat{
		\includegraphics[width=.45\linewidth]{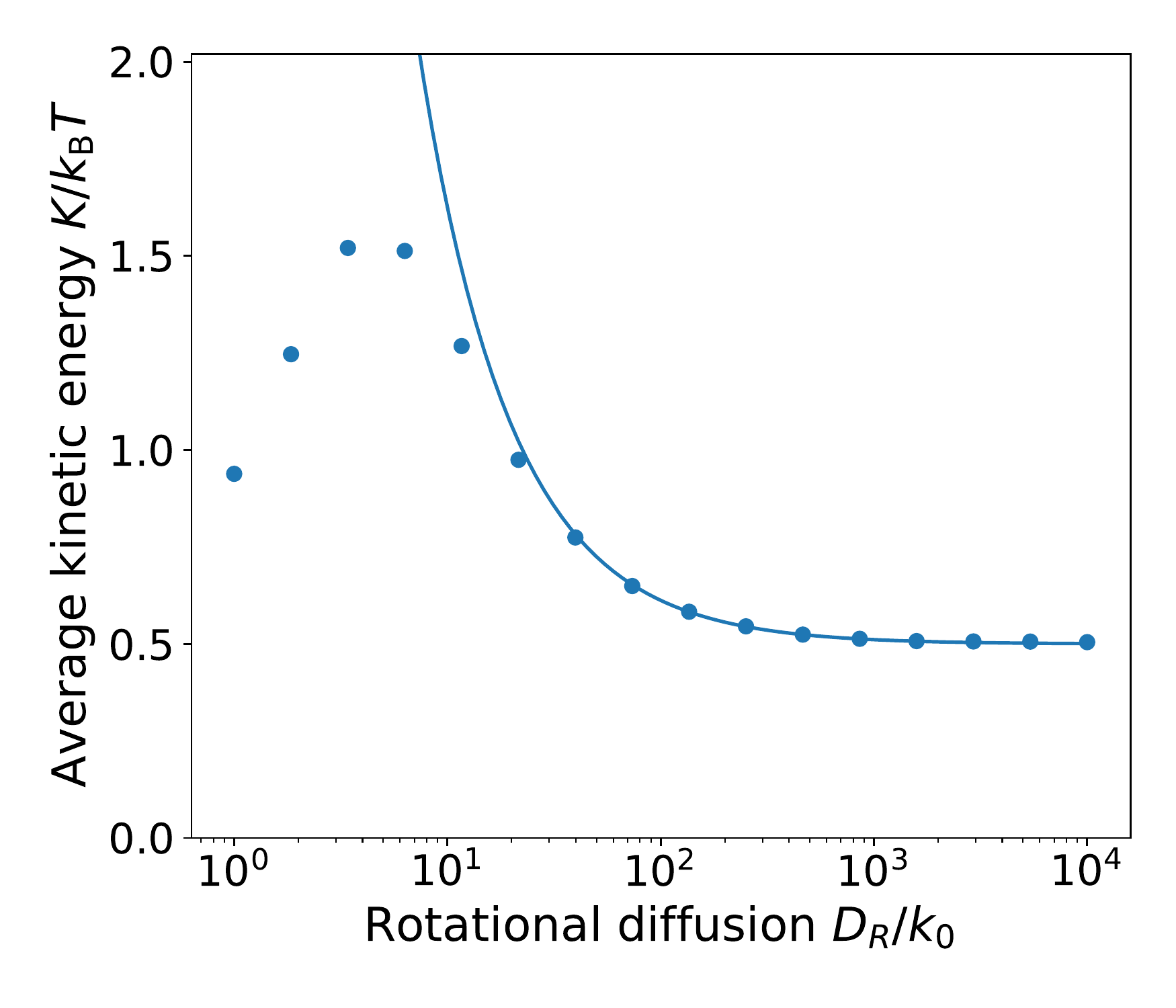}}
    \subfloat{
		\includegraphics[width=.45\linewidth]{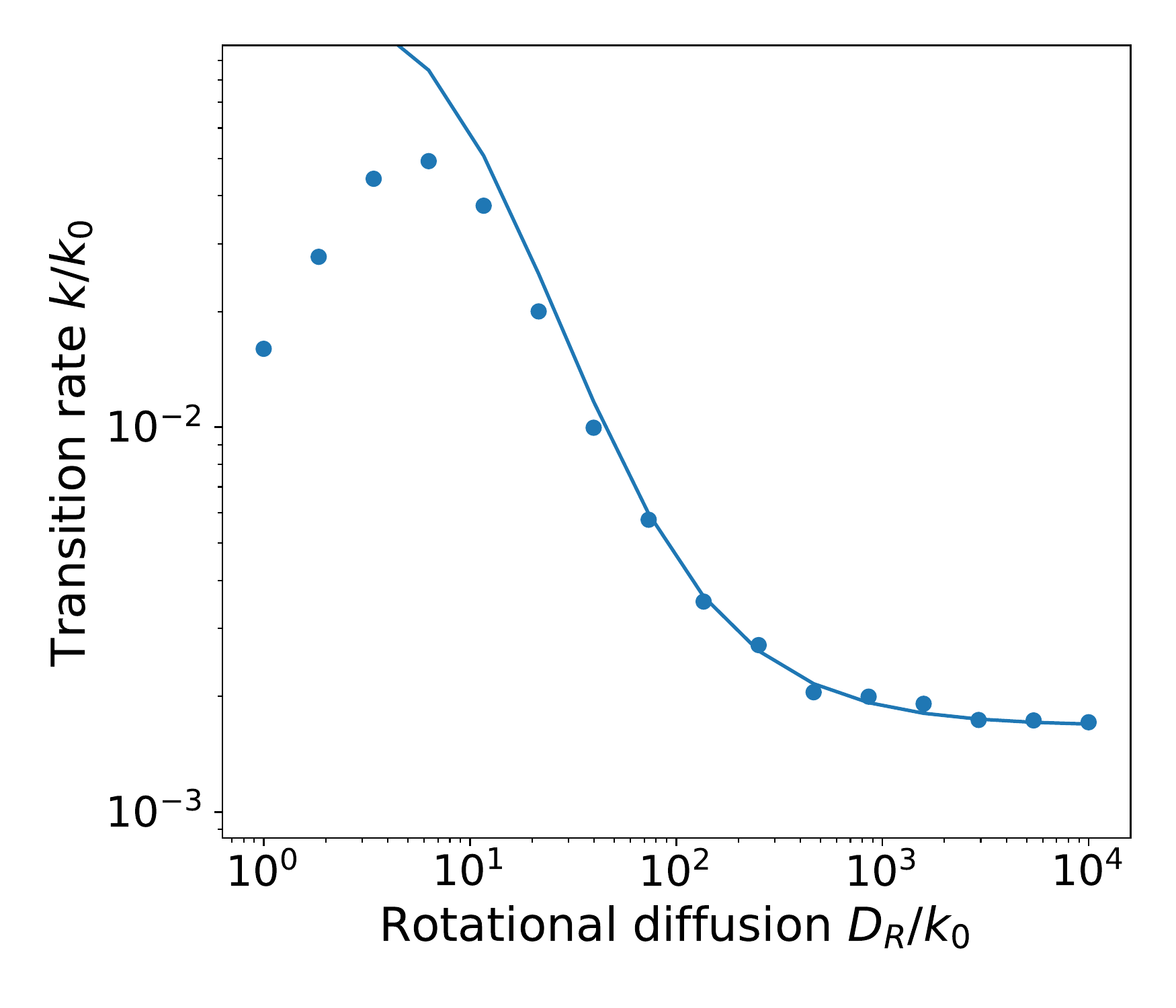}}
	\caption{Average kinetic energy and transition rate constant as a function of the rotational diffusion. The parameters used are the same as those of \figref{\ref{fig:arrhenius}}. The solid lines represent the theoretical approximations of equation~\eqref{eq: effective temperature}, with the approximation for the kinetic energy being parameter-free. The solid curve on the right is fitted to the range of high $D_R$ only, amounting to a vertical displacement on the logarithmic scale. Both theoretical predictions are consistent with the data within their designated regime. Fit parameter: $c_0 =0.648k_0$.}
	\label{fig:hirot}
\end{figure}
The applicability of our findings is highlighted in Fig.~\eqref{fig:hirot}. The average kinetic energy $K$ shown is related to the effective temperature via $K=\kb T\eff /2$ and is approximated extremely well by equation~\eqref{eq: effective temperature}. It remains quite consistent even in close proximity to the active turnover, where the assumptions of our simplification no longer apply. A similar quality is achieved in the parameter-fit showcased by the right panel of Fig.~\eqref{fig:hirot}. The $D_R$ dependence beyond the active turnover seems well-explained by an Arrhenius-like estimate along with the effective temperature provided.

\subsection*{Termination in the Overdamped Regime}

The termination of the active turnover at higher dampings $\Gamma$ shown in \figref{\ref{fig:start}} can be derived in a similar vein to what we discussed in the previous section. The central premise leading to the activity's inclusion in the diffusive term consists of the persistence time being far smaller than the translational timescale. Whether this goal is achieved by speeding up the rotation or inhibiting translational movement is irrelevant. Increasing the friction serves to gradually arrest the dynamics and we end up with the same result as before even without considering the Langevin equation's overdamped limit explicitly. In what follows, the diffusivity still pertains to velocity and not to the position ($D=m\Gamma \kb T$). The expression for the effective temperature
\begin{equation}
	\kb T\eff = \kb T\left(1+\frac{A^2}{2m\Gamma \kb T D_R} \right)
	\label{eq:hiRotRate}
\end{equation}
gives a quantitative estimate about the influence of the active forces on the overall rate landscape at high $\Gamma$, gradually decaying as $A^2/\Gamma$ approaches zero. Irrespective of the activity $A$, one can always select a sufficiently high damping for the active turnover to effectively disappear. On the other hand raising $A$ necessitates a quadratic increase of $\Gamma$ to suppress its effects to the same extent in this approximation. This aspect supports the further expansion of the active turnover into the overdamped regime with increasing $A$ as well as the rate's overall increase, clearly visible in the final panel of \figref{\ref{fig:start}}.

\section{Low Rotational Diffusion Limit}

\subsection*{Infinitely Slow Rotation}

In the extreme case in which the rotational diffusion is infinitesimal, we can consider the angle $\varphi$ as a parameter and the active force as a modification of the potential equal to $-Ax\cos \varphi$. This new, effective potential exhibits different left and right barrier heights that are functions of $\varphi$ \-- we denote them with $\Delta U_L (\varphi)$ and $\Delta U_R(\varphi)$ for the left and right well respectively. Analogously, we define the left and right transition rate as $k_L(\varphi)$ and $k_R(\varphi)$.

\newcommand{\cumu}{\text{cm}}
\newcommand{\intphi}{\int_{0}^{2\pi}d\varphi\;}
\newcommand{\effl}{_{L,\text{eff}\;}}
\newcommand{\effr}{_{R,\text{eff}\;}}
\newcommand{\intx}{\int_{-\infty}^{0} dx\;}
\newcommand{\intv}{\int_{-\infty}^{\infty} dv\;}

First of all, let us introduce an expression for the effective transition rate derived from an ensemble containing all possible modified potentials: For a fixed angle the probability to stay in the left well decays per definition like $e^{-k_L(\varphi)t}$. The long-time decay of a cumulative measure or population that has components individually decaying exponentially is dominated by the slowest rate present. Nonetheless, we need to extract an effective rate that circumvents addressing these individual components directly - a point of view that summarizes all potential tilts into a single rate equation.

To this end, consider the average transition probability $p_{L,\cumu}$ at low waiting times $\Delta t$ (i.e.\ the probability to observe a jump between wells within this duration)
\begin{equation}
p_{L,\cumu} = \int_{0}^{2\pi}d\varphi\; \left( 1- e^{-k_L(\varphi)\Delta t} \right) p_L(\varphi) =: 1- e^{-k\eff \Delta t},
\end{equation}
where $p_L$ denotes the probability density to find the particle in orientation $\varphi$ under the constraint that it currently resides in the left well. 
The integral introduces an effective rate $k\eff$ that generally depends on $\Delta t$. Nonetheless, at $\Delta t$ much smaller than the multiplicative inverse of all rates present, one can conveniently expand the above expression to obtain 
\begin{equation}
\label{eq: ensemble}
k\eff = \intphi k_L(\varphi)p_L(\varphi),
\end{equation}
equalling the standard expectation value. The short time behaviour (i.e., shorter than any rate and of course diffusion timescales) of the left and right well concentrations is then consistently represented by the original ODE as
\begin{equation}
\dot{c}_L = -k\effl c_L + k\effr c_R.
\end{equation}
In the case of infinitely slow rotational diffusion we are in the fortunate position to know a closed form of the marginal angular distribution. Under this condition each angle can be treated as a separate case, allowing us to recycle the result shown in Eq.~\eqref{eq:rateEqSolv}. Up to a normalization factor it reads
\begin{equation}
p_L (\varphi) \propto \dfrac{k_R(\varphi)}{k_L(\varphi)+k_R(\varphi)}.
\end{equation}
In its current form, our result is unfortunately independent of $D_R$. In the next section we introduce a straightforward variant of the above model that amends this shortcoming and discuss the conditions for its applicability.

\subsection*{Reaction-Diffusion Model}

Rate equations are typically not concerned with details about the velocity and position beyond classifying the current state, allowing us to discard spurious details. We can thus integrate the irrelevant variables over the phase space portion corresponding to the state. Proceeding with this line of thought transforms the FPE into a reaction-diffusion equation as shown in the following.

Without loss of generality, we place the boundary between the two wells at $x=0$ and define the left-well population as
\begin{align}
c_L (\varphi, t) := \intv \intx p(x,v,\varphi,t).
\end{align}
We apply the same integrations to the FPE in~\eqref{eq: FPE} to obtain
:
\begin{align}
\label{eq: FPE simplified}
\partial_t c_L  & =D_R \partial_{\varphi\varphi}^2 c_L + \intv \intx \left[  -v\partial_x p + \partial_v(\Gamma v+U'(x)-A\cos\varphi)p+ 
D\partial_{vv}^2 p\right] = \nonumber\\
& = D_R \partial_{\varphi\varphi}^2 c_L - \intv v p (x=0,v,\varphi,t). 
\end{align}
Under the reasonable assumption that the probability distribution decays faster than any linear (typically polynomial) function, all terms containing a derivative of the velocity drop out. A similar idea was applied to the lower limit of the position integral. Alternatively, one can assume a confined/compact initial condition, which generates a time-dependent distribution that vanishes at infinity anyhow. Our expressions are still exact at this point under these conditions. We shall now proceed to apply physically motivated approximations for the remaining integral term of equation~\eqref{eq: FPE simplified}, which describes the net probability flux through the dividing surface. 

We exclude stochastic recrossings where the particle velocity changes its sign in the immediate proximity of the dividing surface. These cover a very small fraction of all reactive paths and contribute practically nothing to the integral for one of two reasons: Either the velocity multiplied to the local distribution is already very small or the sign reversal becomes extremely unlikely. From this standpoint, almost all instances of negative velocities on the barrier come from the right well and vice versa, splitting the integral into two terms associated with the particle origin and thus proportional to the respective local concentration $c_L$ or $c_R$.

In the next step, we use the fact that since the rotational diffusion is small, one can treat the angle as constant at least over the duration of the average transition path time. The probability to visit the barrier top is then approximately proportional to the Boltzmann factor of the effective potential including its linear tilt due to active propulsion $U\eff(x,\varphi) = U(x)-Ax\cos \varphi$. In the context of this conservative-like dynamics, the velocity part of the distribution should factorize as usual and appear in terms of the kinetic energy. The velocity integral thus becomes an additional prefactor, similar to the normalization of the local Boltzmann distribution. Therefore we arrive at 
\begin{align}
\partial_t c_L(\varphi,t)  & =D_R \partial_{\varphi\varphi}^2 c_L -  z_L e^{-\beta \Delta U_{\text{eff},L} }c_L
+z_R e^{-\beta \Delta U_{\text{eff},R} }c_R \nonumber\\
& = D_R \partial_{\varphi\varphi}^2 c_L -  k_L(\varphi) c_L
+k_R(\varphi) c_R. 
\end{align}
The velocity parts and normalizations are absorbed into the dynamic prefactors $z_L,z_R$. In a symmetric potential these quantities become equivalent and once again constitute the only parameter(s) of the model.
Together with the Boltzmann factor we can identify their product as angle-dependent rates, resulting in a reaction diffusion equation much simpler than the original FPE as it depends on only two variables instead of the original four. 

Deriving the analogous expression for the right well and using the constraint of an overall uniform marginal angle distribution in the interval $[0,2\pi)$, we end up with a single equation governing the stationary state
\begin{equation}
D_R \partial_{\varphi\varphi}^2 c_L = [k_L(\varphi)+k_R(\varphi)]c_L  -\frac{1}{2\pi} k_R(\varphi).
\end{equation}  
In presence of the boundary conditions of positivity and periodicity, this ordinary linear differential equation has, in principle, a unique solution. It can even be further simplified to a system of first-order ODEs by defining $w_L = \partial_\varphi c_L$
\begin{equation}
\label{eq: final low DR}
\frac{d}{d\varphi} \left( \begin{array}{c}
c_L\\ w_L \end{array}\right)  = 
\left( \begin{array}{cc}
0 & 1 \\
k_L+k_R & 0
\end{array}
\right) 
\left( \begin{array}{c} c_L\\ w_L \end{array}\right)-
\frac{1}{2\pi}\left( \begin{array}{c}
0\\ k_R \end{array}\right).
\end{equation}

\begin{figure}[tbp]
	\centering
    \subfloat{
		\includegraphics[width=.45\linewidth]{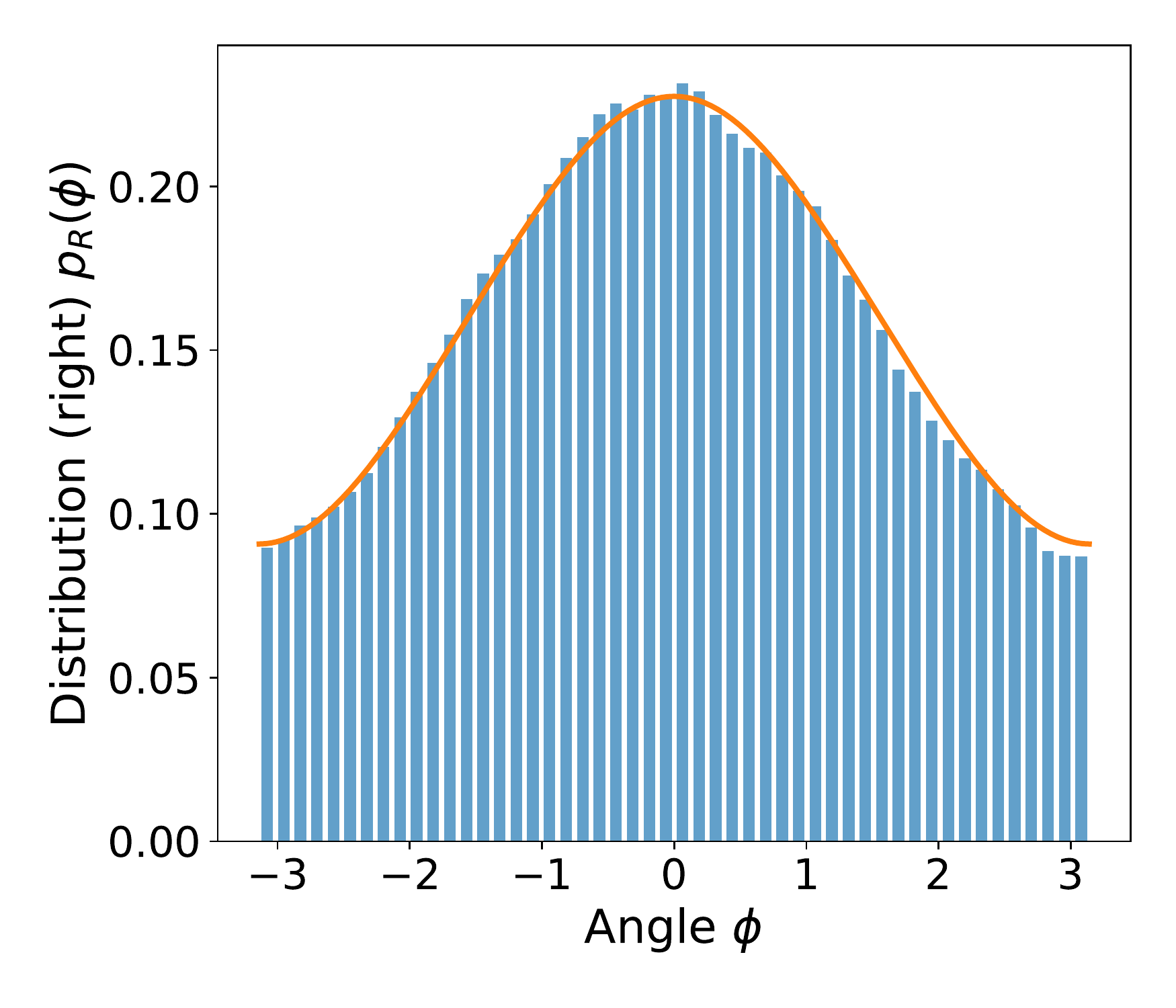}}
    \subfloat{
		\includegraphics[width=.45\linewidth]{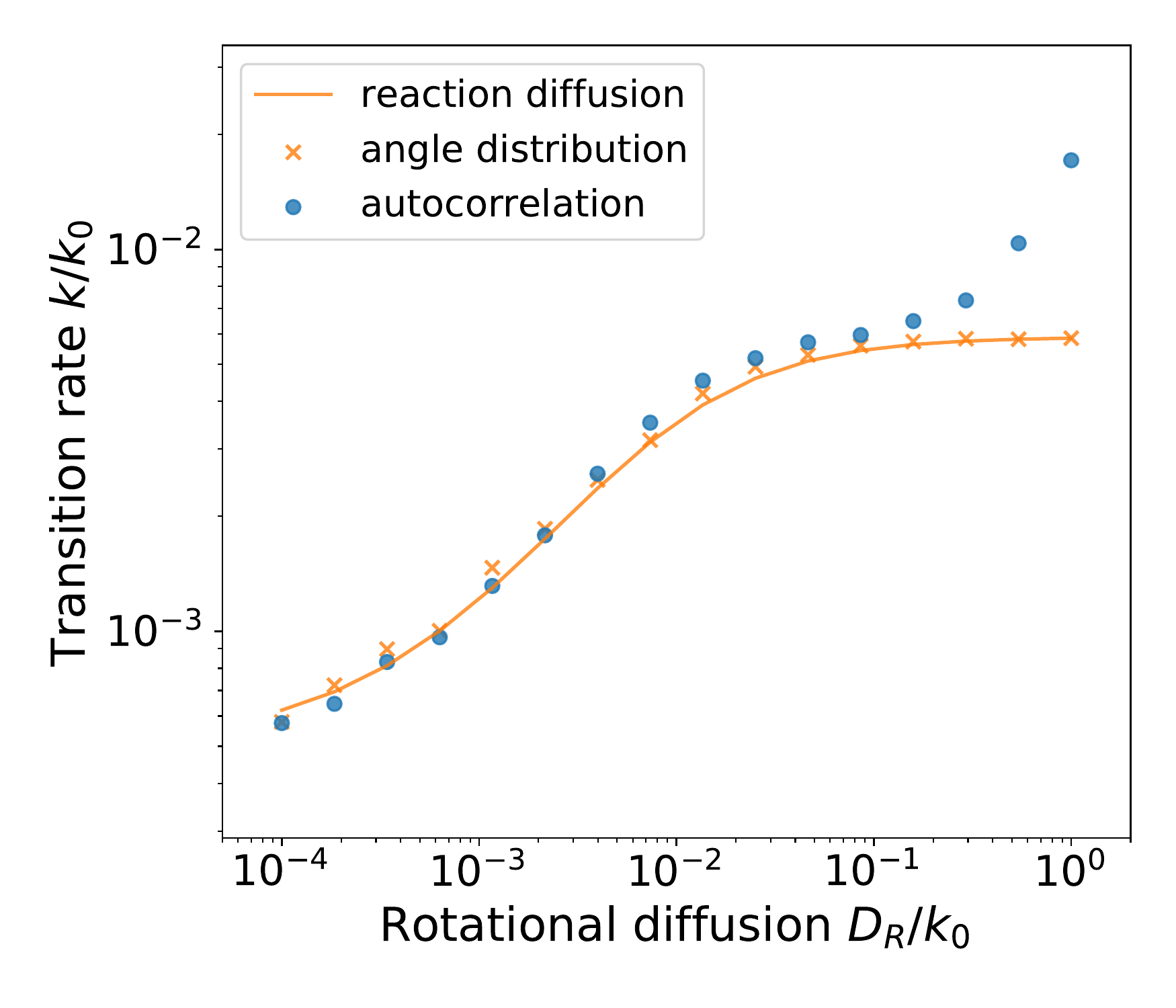}}
	\caption{Left: Example of a $\varphi$ distribution under the condition that the particle sojourns in the right well ($D_R=\num{1.35e-2}k_0$). Right: Comparison of transition rates extracted from the autocorrelation method (blue circles) against Boltzmann-factor averages using the observed distributions (orange crosses). The same simulation parameters were used as in the high $D_R$ investigation, but the time step was increased to obtain longer trajectories at the same expense (each by a factor ten). Additional estimates computed via the reaction diffusion model are included as well, depicted as solid orange lines in the distributions and rate estimates. Using a constant prefactor $z_L=z_R=0.3k_0$ as the only parameter in the model, we attempt to reproduce the real angle distributions and therefrom estimated rates. The assumption of constant prefactors $z_L, z_R$ appears to hold within the regime of validity of the averaging approximation.}
	\label{fig:reacDif}
\end{figure}

For finite (and low) rotational diffusivity, the solution of Eq.~\eqref{eq: final low DR} represents the stationary distribution to calculate the effective rate as seen in Eq.~\eqref{eq: ensemble}.
To reiterate, the central requirement for the validity of our effective model is only a matter of the rotational diffusion's relative smallness and our expressions finally include $D_R$ explicitly. Unfortunately, even this simplified model does not seem to have an easily accessible closed solution due to the complex form of our rate function. The theory's performance can be checked by means of computational data, however, highlighted in Fig.~\eqref{fig:reacDif}. The showcased example of an angle distribution is fully in line with our expectations, with angles favouring transitions into the respective well becoming increasingly more prominent at lower $D_R$.

Furthermore, the right panel depicts the slow rotation part of the rate landscape from last section. Besides increasing the time step and trajectory length by a factor ten, the simulation parameters have been left they as were. The angular distributions along the active modification of the potential are used to compute the rate up to a common prefactor (amounts to a simple displacement in logarithmic scale, the value of the lowest point is made to coincide with the autocorrelation data). Our predictions turn out to be very accurate for an extensive regime of rotational diffusivity values and capture the rate's initial increase up until the theory reaches its high $D_R$ plateau. At that point we are faced with a situation where our initial premises and conditions fall apart, making this discrepancy fully expected. The theory seems like an adequate approximation up until one to two orders of magnitude before reaching the active turnover, falling short of the fast rotation limit in terms of accurate coverage.

\begin{figure}[tbp]
	\centering
	\includegraphics[width=0.6\linewidth]{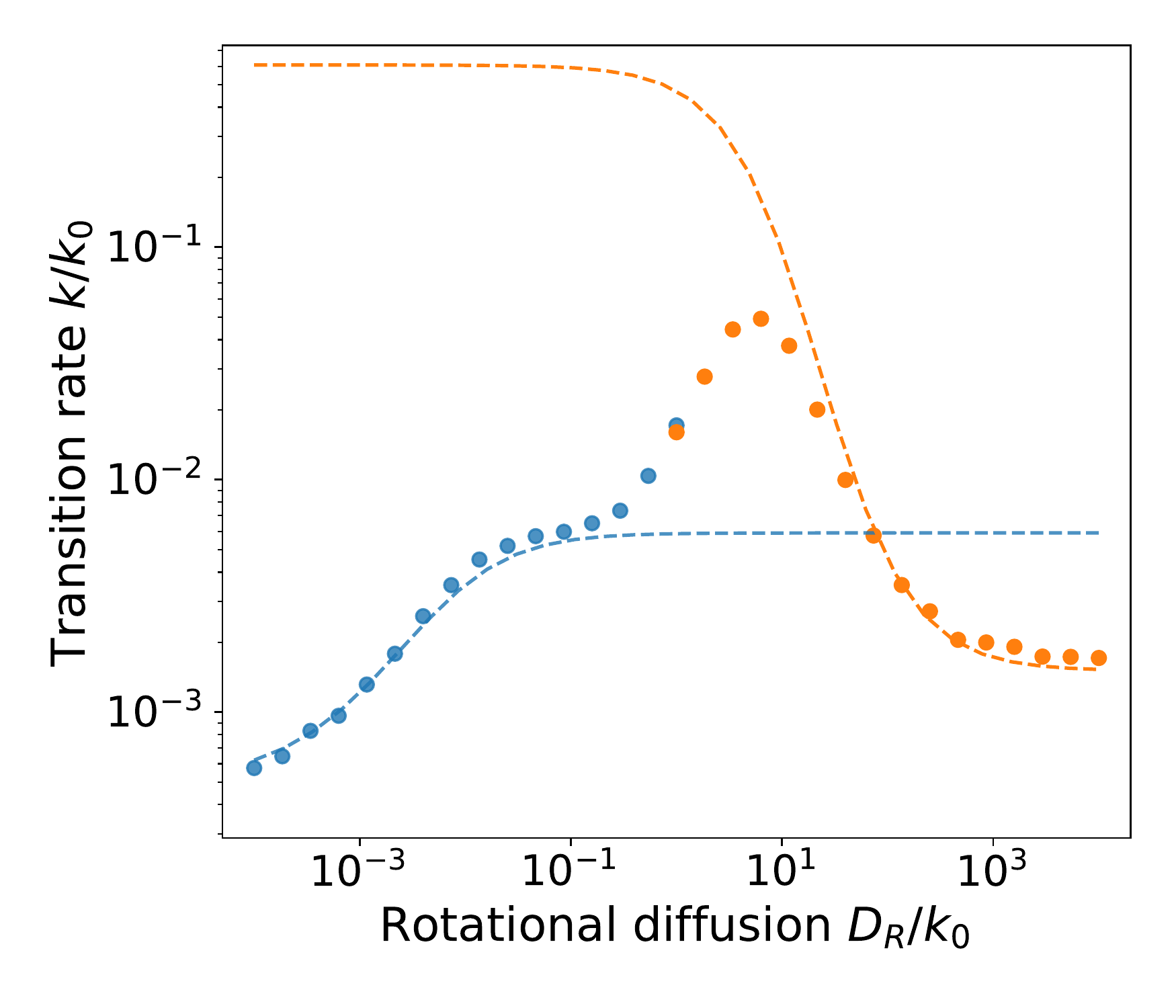}
	\caption{Side-by-side depiction of the high and low $D_R$ data and plots. Outside the regime of the turnover itself, our models make very accurate predictions about the landscape's shape, capturing its key features.}
	\label{fig:doubleLim}
\end{figure}

Finally, \figref{\ref{fig:doubleLim}} showcases the results of both models alongside another, accurately replicating the landscape over all possible $D_R$ safe for the immediate proximity of the active turnover. Both approaches only rely on the dynamic prefactor as parameter, corresponding to simple displacements of the curves on a logarithmic scale. The good consistency between theory and data hints at both theories capturing the essential details affecting the rate.

\section{Notes on the Simulation}
\label{sec:simu}

In this section, we summarize the general outline of the simulations conducted. We propagate trajectories in one spatial dimension by using the OVRVO integrator first devised by Sivak et al.\ \cite{Sivak2014b} and based on the well-known Liouville framework for symplectic integrators. Compared to a standard velocity Verlet integrator, this method additionally encompasses appropriate Ornstein-Uhlenbeck steps placed at the beginning and the end of each iteration to act as the system's thermostat. Each data point in the transition landscapes results from a trajectory of duration $T$ generated with time step $\Delta t$, specified in the description of the respective figures.  
As all rate coefficients shown were extracted from long trajectories and reflect the stationary state, the initial condition becomes irrelevant. For completeness, however, let us note we generally start in one of the minima with a velocity drawn from the Maxwell distribution. 
Next, we proceed to delineate the analysis methods applied to both numerical and experimental trajectories.

\begin{figure}[tbp]
	\centering
	\subfloat{
		\includegraphics[width=.45\linewidth]{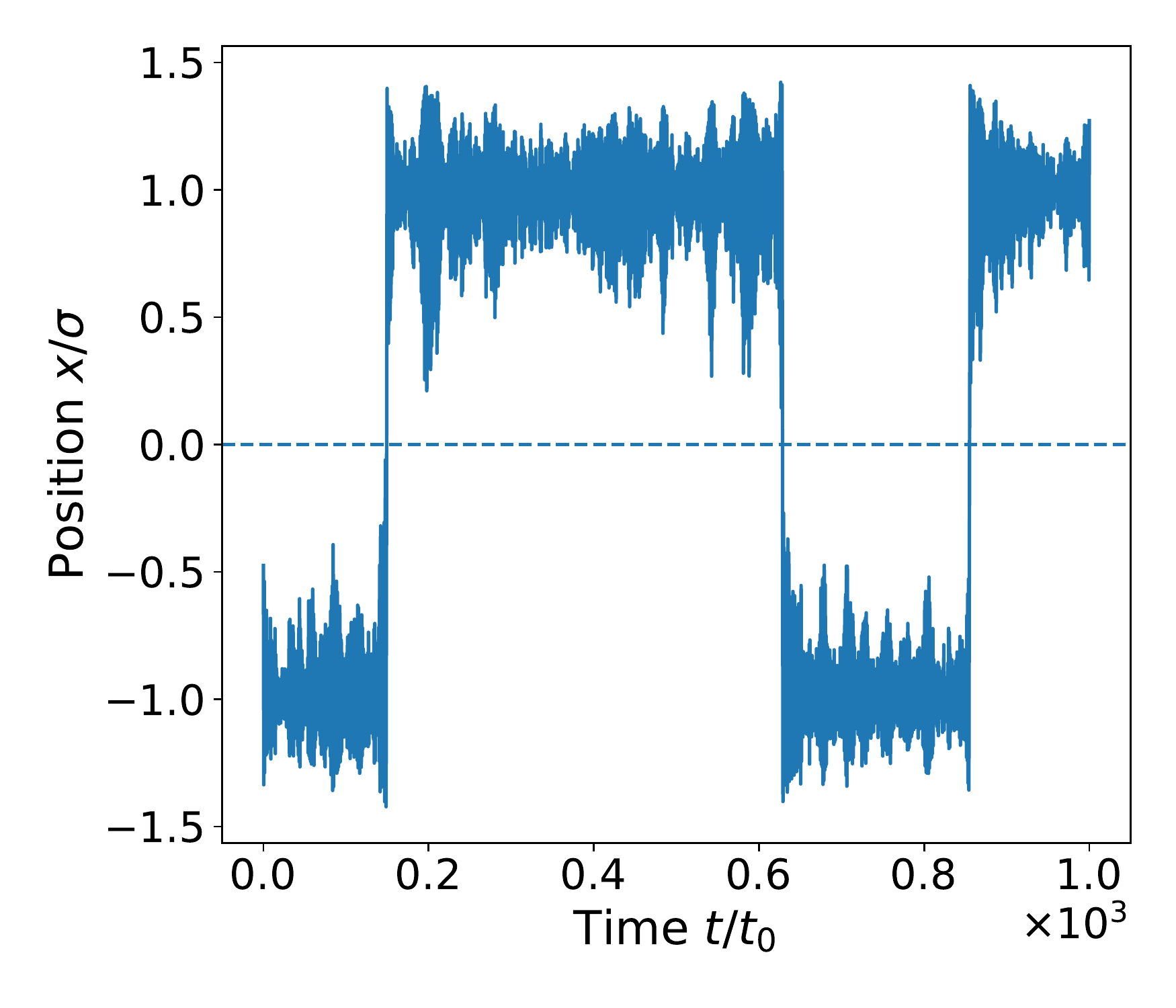}}
    \subfloat{
		\includegraphics[width=.45\linewidth]{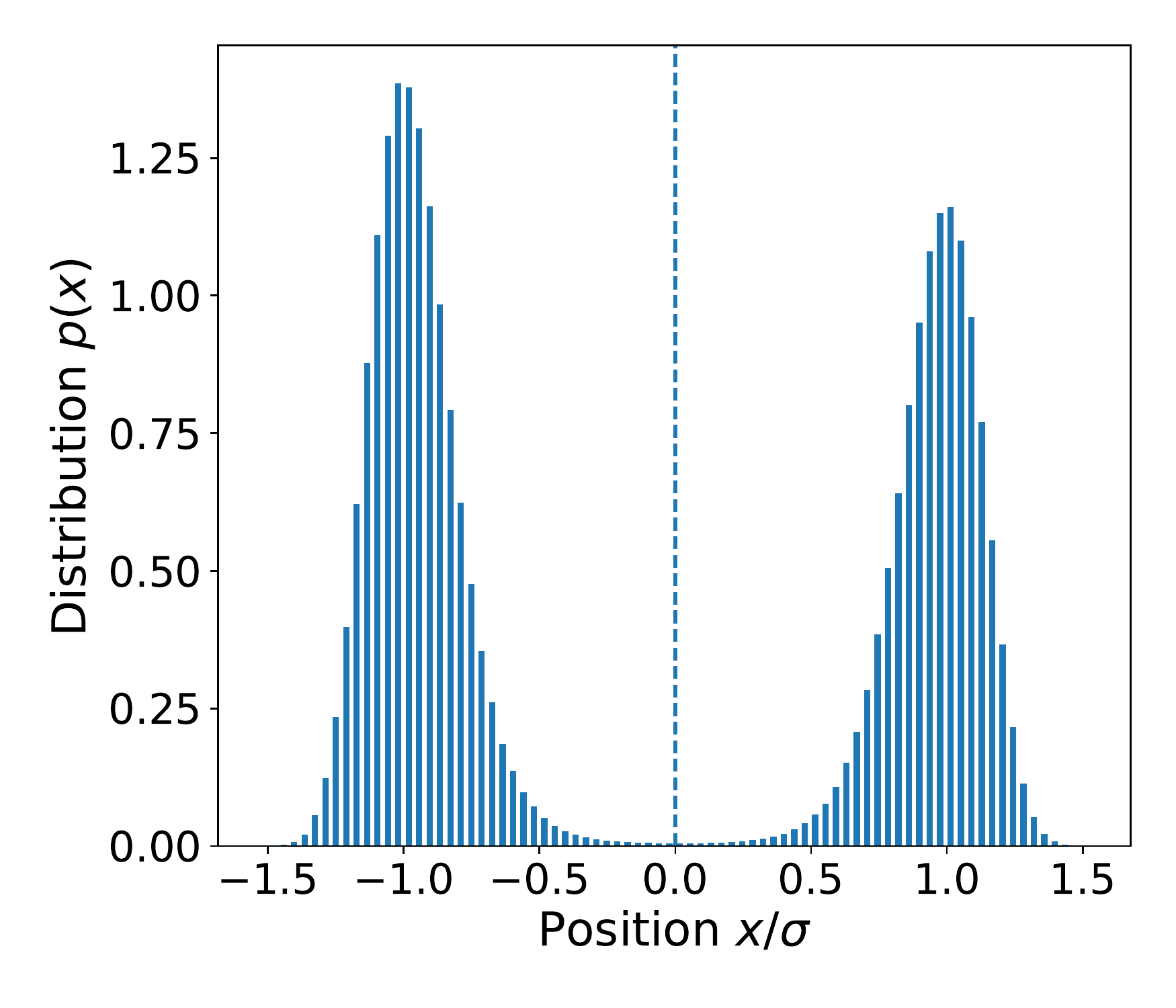}}
	\caption{Segment of a trajectory in a bistable trap (left) and corresponding full-length histogram (right), belonging to the data shown in \figref{\ref{fig:hirot}} at $D_R=\num{4.64e+2}k_0$. The dashed line demonstrates the dividing surface's location.}
	\label{fig:surface}
\end{figure}

The first step of any rate extraction algorithm consists in the selection of an appropriate dividing surface. Let us direct our attention to Fig.~\eqref{fig:surface}, showcasing a segment of a simulated trajectory. The peaks in the distribution corresponding to the metastable states are separated by a range of low probability. For signals of this kind, the extracted rate does not depend sensibly on the dividing surface's exact location. Shifting it slightly to the left or right influences our results only negligibly, which is why a straightforward selection method fully serves our purposes. We locate the approximate position of the maxima left and right by generating histograms with a prescribed bin width and place our surface in the minimum between those points. This approach relies on sufficiently smooth histograms to yield appropriate results, which all investigated trajectories (experimental or computational ones) were able to provide.

\begin{figure}[tbp]
	\centering
	\includegraphics[width=0.5\linewidth]{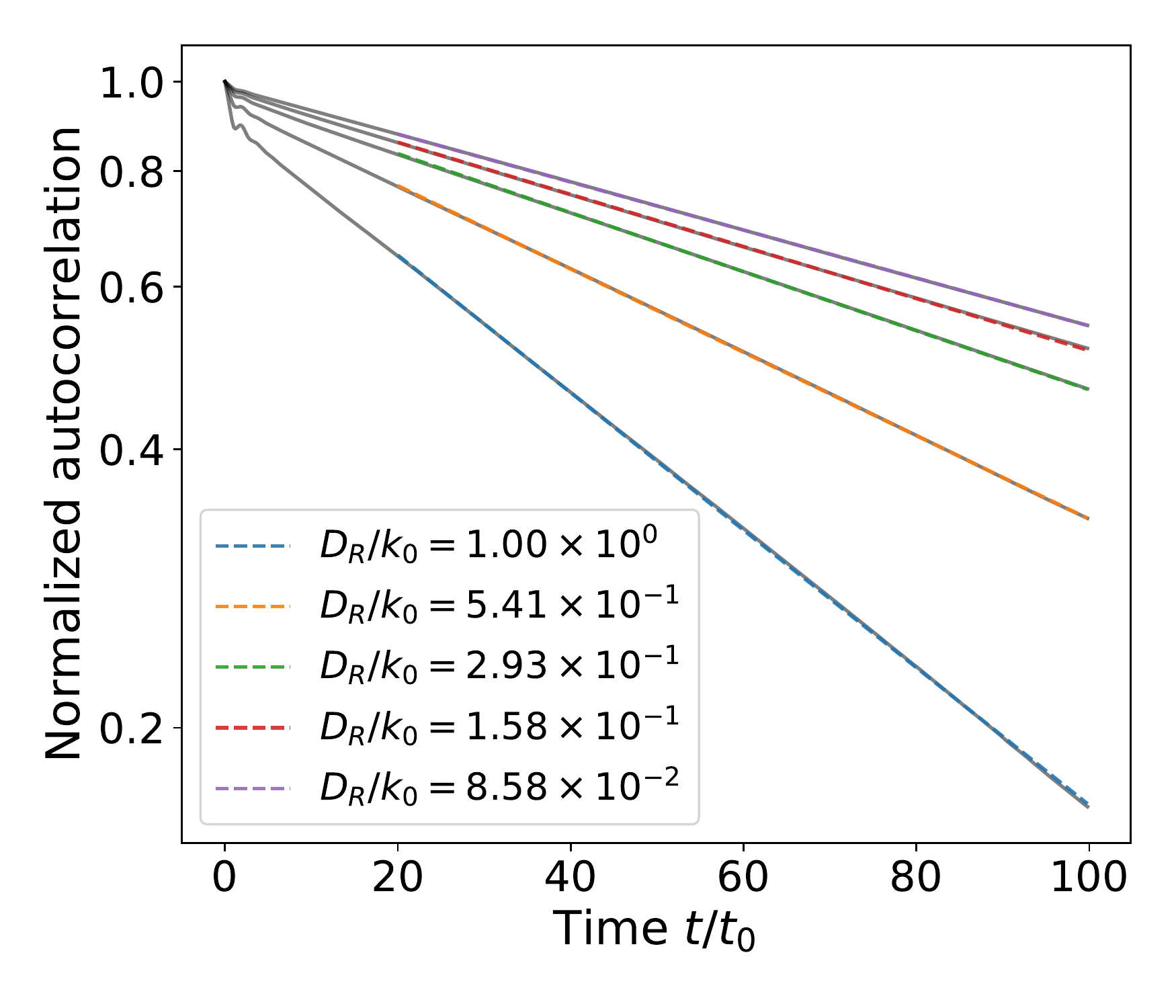}
	\caption{Normalized autocorrelation functions for five  values of the rotational diffusion. The fits (dashed lines) exhibit an excellent agreement with the computational data (gray lines). At very short timescales deviations from the predicted exponential decay occur, which are excluded from the fitting regime.}
	\label{fig:autoLog}
\end{figure}

Having placed the dividing surface, our signal is converted into an indicator function to subsequently evaluate its normalized autocorrelation, an example of which is shown in Fig.~\eqref{fig:autoLog}. We invoke the Wiener-Khinchin theorem and compute all autocorrelations via FFTs for computational efficiency.
The initial behaviour of our autocorrelations is very well-represented by a simple exponential decay except a very brief period at the very beginning attributed to short-time correlations. The curves' roughness increases as the autocorrelation approaches zero as a consequence of fewer uncorrelated indicator pairs, but can be diminished by analyzing longer trajectories (i.e., gathering more statistics). Excluding the initial non-exponential deviation, we fit an exponential of the form $c_0 e^{-kt}$ to these curves, limiting the regime to the part of the normalized autocorrelation larger than a prescribed threshold, here $0.10$.
The parameter $k$ should then correspond to the sum of left and right transition rates, completing the extraction.

Uncertainties have only been computed for the data presented in the main text, evaluated by splitting the trajectories in ten segments each. We extract the rate for each segment and compute the standard deviation of the mean results as our error estimates.

\subsection*{Replication Study}

In this section we describe how we computationally replicated the experimental findings seen in the main text, specifically the rate landscape visualized in Fig.~3. Such an endeavour provides additional insight by unearthing differences and physical contributions not captured by the simple Langevin model. Since the signal mainly serves to distinguish between wells rather than providing an accurate position at any point of time, it is not inherently well-suited to infer the experimental potential. Unable to directly extract the shape of the physical potential from the signals, we opt for a model that allows to prescribe all well-known information along with the remaining parameters to be optimized.

The key quantities of interest are the approximate widths of the two wells and of the barrier (or alternatively their frequencies) to account for entropic effects, the barrier height $ \Delta U$, and the 
activity $A$. The two well-frequencies have been measured experimentally, amounting to $\omega_L = 2\pi \times (73.0\pm 0.5)\SI{}{kHz}$ and $\omega_R = 2\pi \times (82\pm 4)\SI{}{kHz}$. The remaining three parameters, i.e.\ activity, well-separation, and barrier height are tuned to find the best replication of the experimental landscape as presented in the main text (Fig.~3). Initial guesses for the activity and barrier height have been extracted directly from the rate landscape by means of parametric fits. The rate approximation in the regime of high rotational diffusion directly contains both $A$ as well as $\Delta U$ within the expression for the expected rate Eq.~\eqref{eq:hiRotRate}. As the experimental data does not indicate a significant difference between left and right barrier height, we assume them to be equal. 

As for the conservative trapping/barrier force we employ a piecewise parabolic potential consisting of three segments, straightforwardly allowing us to prescribe the well frequencies:
\begin{equation}
	U(x)=
	\begin{cases}
	\dfrac{m\omega_L^2}{2}  (x-a_L)^2 &
	\quad \text{for } x<b_L,  \\[10pt]
	-\dfrac{m\omega_B^2}{2} x^2 + \Delta U &
	\quad \text{for } b_L\leq x\leq b_R, \\[10pt]
	\dfrac{m\omega_R^2}{2} (x-a_R)^2 &
	\quad \text{for } x>b_R.
	\end{cases}
\end{equation}
Imposing the aforementioned parameters and additionally demanding continuity of the potential and its first derivative fully determines the location of minima $a_L, a_R$ and connection points between segments $b_L, b_R$. The expressions for $a_L, b_L$ read
\begin{align}
	a_L &= -\sqrt{\frac{2\Delta U(\omega_L^2+\omega_B^2)}{m\omega_B^2 \omega_L^2 }}, \nonumber\\
	b_L &= \frac{\omega_L^2 a_L}{\omega_B^2+\omega_L^2}.
\end{align}
The expressions for the right side are obtained by replacing the respective index and changing the sign for $a_R$.

As this approximate one-dimensional potential represents a simplification of the experimental system, we aim to locate a point in parameter space that generates a rate landscape as consistent as possible in selected key aspects. The key aspects considered include the general magnitude of rates, the height, position and width (at half height) of the passive and active turnovers on a logarithmic scale. To this end we apply a bisection-like approach on a discrete grid of parameter values. We consider $A$, $\omega_B$, and $\Delta U$ in steps of $\SI{0.05}{fN}$, $0.05 \omega_L$, and $0.01 \kb T$, respectively. 
Since the Kramers turnover in the regime of fast rotation is largely independent of $A$, the respective experimental data can be used to infer the barrier height and frequency/well-separation separately. Leaving the range of potential values more liberty than would be necessary from these initial estimates, we limit the search intervals to $\Delta U/\kb T \in [1.9,2.3]$ and $\omega_B/\omega_L \in [0.5,2.]$. The permissible activity values are restricted by the experimental estimate in conjunction with its symmetric uncertainty. We find excellent agreement between simulation and experiment in terms of turnover heights and widths with the parameter set $A = \SI{9.35}{fN}$, $\omega_B=0.60\omega_L$ (translating into a minima separation of $\SI{0.66}{\mu m}$), and $\Delta U = 2.065\kb T$. 

The position of the calculated Kramers turnover coincides with the experimental result within its resolution and uncertainties as well, leaving the position of the active turnover in terms of $D_R$ as the only clear discrepancy despite the ad hoc nature of the potential's functional form. The ratio of the simulated and the experimental position of the turnover on the $D_R$-axis amounts to approximately $0.447$, with the simulation placing it at a lower diffusivity. 
Simulations performed in- and outside the delineated optimization range do not hint at the possibility of situating the two lines of maxima at their targeted positions simultaneously. It appears that the ratio between the active and passive turnover positions remains roughly constant over an extensive regime of parameter values. A possible explanation of this phenomenon lies in both being closely tied to the average transition path time. The Kramers turnover occurs when the particle loses energy on the scale of one $\kb T$ throughout a transition. Similarly, the active turnover stems from the propulsion decorrelating over the duration of a typical transition. Both turnover locations are thus expected to be inversely proportional to the average transition path time in good approximation. 
In one dimension it becomes therefore difficult to change the position of one maximum independently from the other.
Multidimensional potential landscapes, on the other and, offer additional freedom by allowing for a multitude of transition paths that differ already in position space. Transitions induced purely by thermal fluctuations as captured and collected in the Kramers phenomenology typically look for paths crossing the lowest point in the barrier as preferred transition channels, even if this point does not lie on a straight line with the minima. On the contrary, activity-induced transitions events are likely to follow a more "direct" approach, offsetting an increased barrier height in proportion to the magnitude of $A$. 
The different channels associated with active and passive transitions generally translate in distinct transition path times and thus location of the turnover maxima. As the experiment naturally unfolds in a three-dimensional setup, and not all coupling effects between dimensions can be erased in practice, it seems sensible to assume that the lowest energy paths would neither follow a straight line nor perfectly coincide with the active force axis. Any small misalignments of the active force, well- and barrier minima, major axes of the trapping forces in their parabolic approximation etc.\ could induce this effect. 

Finally, it is important to once again underline the consistency of the simulation results with the experimental rates despite the usage of an ad hoc potential. Even the most strongly deviating property, the active turnover's position, only does so by a factor of $c=0.447$, residing in the same order of magnitude. Given a rotational diffusion axis rescaled by this calibration factor, the obtained replicated landscape becomes quantitatively consistent with the experiment everywhere, supporting the models discussed in this document.


\section{Active Noise}

\subsection*{Statistical properties}
\label{sec: activity}

In a reduced phase space consisting of only position $x$ and velocity $v$ (but not the angle $\varphi$), active propulsion is treated as a source of colored noise with persistence time $\tau_A$. In this section, we derive the most central properties of the active force $n(t) = A\cos(\varphi(t))$ from equation~(1). These also serve as consistency tests for the experimentally implemented noise source.

We start by computing the stationary distribution $p(n)$ of $n(t)$. Noting that $\varphi$ evolves randomly according to Brownian motion on the interval $[0,2\pi)$ with periodic boundary conditions, the stationary distribution of $\phi$ must be uniform, $p(\varphi) = 1/(2\pi)$. Applying a change of variable from $\varphi$ to $n$ yields 
\begin{equation}
    p(n) = \intphi \frac{1}{2\pi} \delta(n-A\cos\varphi) = \frac{1}{A\pi |\sin\varphi(n)| } = \frac{1}{\pi \sqrt{A^2-n^2} }.
    \label{eq:activityDistr}
\end{equation}

Next, we direct our attention to the autocorrelation $R_{nn}(t)$ and to the power spectral density $S_{nn}(\omega)$. As the angle evolution represents a Wiener process rescaled by the prefactor $\sqrt{2D_R}$, increments of the angle throughout a time step $\Delta t$ are independently distributed according to a Gaussian with variance $D_R\Delta t$
\begin{equation}
\label{eq: characteristic}
    \Delta \varphi = \varphi(t+\Delta t)-\varphi(t) \propto \frac{1}{\sqrt{2\pi \Delta t D_R }} e^{-\frac{\Delta \varphi^2}{4\Delta t D_R}}.
\end{equation}
In equation~\eqref{eq: characteristic}, the periodicity of $\varphi$ has not been taken into account yet. We impose without loss of generality $\Delta t\geq 0$. For any real stochastic process, we recover the negative time intervals from the property $R_{nn}(\Delta t)=R_{nn}(-\Delta t)$. The activity's autocorrelation can be evaluated as
\begin{align}
    R_{nn}(\Delta t) := \langle n(t+\Delta t) n(t)\rangle  = A^2\intphi \int_{-\infty}^\infty d\Delta \varphi \; \cos(\varphi) \cos(\varphi+\Delta \varphi) p(\varphi) p(\Delta \varphi).
\end{align}
Applying Euler's relation $\cos x = (e^{ix}+e^{-ix})/2$ and 
\begin{align*}
    \left\langle e^{i\varphi}\right\rangle = 0, \qquad
    \left\langle e^{i\Delta \varphi}\right\rangle = e^{-D_R \Delta t},
\end{align*}
leads us to an exact expression for the autocorrelation
\begin{align}
    R_{nn}(\Delta t) = \frac{A^2}{2} e^{-D_R|\Delta t|}.
\end{align}
Consequently invoking the Wiener-Khintchin theorem  we obtain the power spectral density $S_{nn}(\omega)$ (the shorthand $c.c.$ denotes the complex conjugate):
\begin{equation}
\label{eq: Snn}
S_{nn}(\omega) = \frac{1}{2\pi} \int_\mathbb{R} \text{d}t\ R_{nn}(t) e^{-i\omega t} = \frac{A^2}{2\pi}\int_{\mathbb{R}^+} \text{d}t\ e^{-\left(D_R + i\omega\right)t} + c.c. = \frac{A^2}{\pi} \frac{D_R}{\omega^2 + D_R^2}.
\end{equation}

\subsection*{Experimental Characterization}

Figure~2b in the main text depicts a histogram of the active force signal at the output of the FPGA. The output range of the FPGA is $[\SI{-1}{V}, \SI{1}{V}]$, meaning that $A=\SI{1}{V}$. The histogram in Fig.~2b agrees very well with equation~\eqref{eq:activityDistr} for $A=1$. Figure~2c shows the power spectral density $S_{nn}$ of the active force for three different values of the rotational diffusion $D_R$. We fit $S_{nn}$ with Eq.~\eqref{eq: Snn} and extract the cutoff frequency $f_c$. The cutoff frequency allows us to calibrate the physical value of the rotational diffusion using $D_R = 2\pi\times f_c$. The value of $A$ measured in Newton is estimated from the position response of a particle to a known modulated voltage. The experimental value of $A$ equals $\SI[separate-uncertainty = true]{6.8 \pm 3.4}{fN}$. The uncertainty in the estimation of the activity is limited by the uncertainty in the calibration factor of the position measurement \cite{Hebestreit2018}. Specifically, in order to determine a force in Newton, we make use of the mass-dependent transfer function of the particle. An uncertainty up to $50\%$ in the mass of the particle, the main precision bottleneck of the position calibration, translates into the same uncertainty for the estimation of the activity. Note, however, that the estimate of $A/m$ does not suffer from such uncertainty.

\bibliographystyle{apsrev4-1} 
\bibliography{referencesShort} 